\documentclass[11pt]{article}
\usepackage{amssymb,epsf,graphicx}
\newcommand{\be}{\begin{equation}}
\newcommand{\ee}{\end{equation}}
\newcommand{\bea}{\begin{eqnarray}}
\newcommand{\eea}{\end{eqnarray}}
\newcommand{\bdm}{\begin{displaymath}}
\newcommand{\edm}{\end{displaymath}}

\newcommand{\nonu}{\nonumber \\}

\def\ei{\hat{\epsilon}_{(-\frac{1}{2})}}
\def\eii{\hat{\epsilon}_{(\frac{1}{2})}}
\def\a{\alpha}
\def\b{\beta}

\def\d{\delta}
\def\dil{\d_{D}}
\def\e{\epsilon}           % Also, \varepsilon
\def\bare{\bar{\epsilon}}
\def\dage{\epsilon^{\dag}}
\def\Dsl{D \hskip-.6em \raise1pt\hbox{$ / $} }
\def\f{\phi}               %       \varphi
  
\def\g{\gamma}
\def\h{\eta}
\def\half{\frac{1}{2}}
\def\hata{\hat{a}}
\def\hatb{\hat{b}}

\def\hati{\hat{i}}
\def\hatj{\hat{j}}

\def\hatn{\hat{\nu}}
\def\hatr{\hat{r}}
\def\hatrr{\hat{r}}
\def\hatt{\hat{t}}

\def\h1{\hspace{1cm}}

\def\inf{\infty}

\def\k{\kappa}             % Also, \varkappa (see below)
\def\l{\lambda}
\def\m{\mu}
\def\mhalf{-\frac{1}{2}}
\def\n{\nu}
\def\na{\nabla}
\def\o{\omega}  
\def\p{\pi}                % Also, \varpi
  \def\th{\theta}                  %     \vartheta
\def\r{\rho}                                     %     \varrho
\def\s{\sigma}                                   %     \varsigma
\def\t{\tau}
\def\th{\theta}
\def\til{\tilde}

\def\x{\xi}

\def\z{\zeta}

\def\G{\Gamma}
\def\Gt{\Gamma^{\hat{t}}}
\def\Ga{\Gamma^{\hat{a}}}

\def\Gr{\Gamma^{\hat{r}}}

\def\L{\Lambda}
\def\O{\Omega}

\def\Pm{P^{-}}
\def\Pp{P^{+}}
\def\pa{\partial}

\def\S{\Sigma}

\def\co{{\cal O}}
\def\calO{{\cal{O}}}

\def\pa{\partial}

\newcommand{\Tr}{\mathop{\rm Tr}\nolimits}

\def \tG {{}^tG}

\def\sla#1{\rlap{\hbox{$\mskip 1 mu /$}}#1}% good slash for l.c.
\newcommand{\sd}{\sla \nabla}

\begin{document}

\begin{titlepage}
\begin{flushright}
ITFA-2005-20 \\
hep-th/0506123
\end{flushright}
\vfill
\begin{center}
{\LARGE\bf Positivity of energy for asymptotically
locally AdS spacetimes}
\\ \vskip 10.mm {\Large  Miranda C.N. Cheng
and Kostas Skenderis } \\ \vskip 7mm {Instituut voor Theoretische
Fysica, Universiteit van Amsterdam,
Valckenierstraat 65,
1018 XE Amsterdam,
The Netherlands}
\end{center}
\vfill
\begin{center}
{\bf ABSTRACT}
\end{center}
We derive necessary conditions for the spinorial Witten-Nester energy to be
well-defined for asymptotically locally AdS spacetimes.
We find that the conformal boundary should admit a spinor satisfying 
certain differential conditions and in odd dimensions
the boundary metric should be conformally Einstein. 
We show that these conditions 
are satisfied by asymptotically AdS spacetimes. The gravitational
energy (obtained using the holographic stress energy tensor) and 
the spinorial energy are equal in even dimensions
and differ by a bounded quantity related to the conformal anomaly 
in odd dimensions.
\vskip 7mm
\vfill
\hrule width 5.cm
\vskip 2.mm
{\small
\noindent {\tt mcheng,skenderi@science.uva.nl}}
%\end{quote}
%\begin{flushleft}
%PACS 11.25.-w, 04.65.+e
%\end{flushleft}
\end{titlepage}

\tableofcontents
\addtocontents{toc}{\protect\setcounter{tocdepth}{2}}

\newpage

\section{Introduction}

The AdS/CFT duality relates gravity in asymptotically AdS spacetimes
to a quantum field theory on its conformal boundary. One of 
the main features of the duality is that the boundary fields parametrizing 
the boundary conditions of bulk fields are identified
with QFT sources that couple to gauge invariant operators. 
In particular, the boundary 
metric $g_{(0)}$ is considered as a source of 
the boundary energy momentum tensor and
at the same time is the metric of the 
spacetime on which the dual field theory is defined. 

In quantum field theory the sources are unconstrained, 
so that one can functionally differentiate w.r.t. them to 
obtain correlation functions. Thus the duality 
requires the existence of spacetimes associated 
with general Dirichlet boundary conditions for the metric.
Such general boundary conditions go beyond what has been
considered in the GR literature
where asymptotically AdS spacetimes (AAdS) were defined to
have a specific asymptotic conformal structure, namely that of  
the exact AdS solution \cite{AM,Henneaux:1984xu}, 
but they have been considered in the mathematics literature \cite{FG}.
The asymptotic structure of these  more general spacetimes 
is only locally that of AdS; we will call them
asymptotically locally AdS (AlAdS) spacetimes. 
 
An integral part of the correspondence is how 
conserved charges are mapped from one side to the 
other. This is in fact dictated by the basic
AdS/CFT dictionary. On the field theory
side, conserved charges are generated by 
conserved currents. In particular, the energy
can be computed from the energy momentum tensor.
Applying the AdS/CFT dictionary, we find that 
one should be able to compute 
the gravitational energy from the energy momentum tensor  
obtained by varying the on-shell gravitational
action w.r.t. the boundary metric. Such a
definition of conserved charges is available in the literature
\cite{Brown:1992br}, but the naive implementation of this idea
gives infinite answers, essentially due to the 
infinite volume of spacetime. In the GR 
literature such infinities are (explicitly or implicitly) 
dealt with by background subtraction.
The AdS/CFT correspondence, however, suggests a new
approach: one subtracts the infinities by means of
boundary counterterms \cite{HS}-\cite{PS3},
as done in QFT in the process of renormalization. 
This procedure, called holographic renormalization, 
is by now a well studied method.
We will call the charges defined 
using the holographic energy momentum tensor 
``holographic charges''. Notice that these
charges are defined intrinsically 
rather than relative to some other spacetime. This is a definite
advance over the background subtraction method, since a suitable 
reference spacetime does not exist in general.

The holographic charges agree\footnote{The apparent difference between
the holographic mass for odd dimensional AAdS and other definitions
such as the one in \cite{AM} is now understood to be due to the 
fact that these other approaches effectively compute masses
relative to that 
of $AdS_{2k+1}$, and the holographic mass of $AdS_{2k+1}$
is nonzero, see \cite{PS3}
for a detailed discussion.} with previous definitions
of conserved charges
\cite{Abbott:1981ff,AM,Henneaux:1984xu,Hawking:1995fd} 
when the latter are applicable, i.e. when the spacetime
approaches that of the exact AdS solution and one 
considers the energy relative to AdS, see 
\cite{Hollands:2005wt} for a detailed comparison between different
definitions of conserved charges. The new definition on the other hand 
extends to arbitrary asymptotically locally AdS spacetimes. 
Moreover it was proved in \cite{PS3} that these charges 
arise as Noether charges associated to asymptotic
symmetries of such spacetimes and also shown to agree with 
the charges defined in the
covariant phase space approach of Wald {\it et al} 
\cite{Wald&Zoupas}.

AAdS spacetimes are known to have positive mass 
relative to the AdS solution, which saturates 
the bound in a positive mass theorem \cite{GHW}.
The proof in \cite{GHW} generalizes Witten's 
spinorial positive energy theorem \cite{Witten:1981mf}
for asymptotically flat spacetimes. A natural question to ask 
is whether the holographic energy of $AlAdS$ spacetimes 
is subjected to a positivity theorem. Such a generalization
is far from obvious and is known to be false for
asymptotically locally flat spacetimes \cite{Witten:1981gj,lebrun}.   
A new positivity theorem for a specific class 
of AlAdS spacetimes has been conjectured in 
\cite{Horowitz:1998ha}.
In this reference an $AlAdS$ solution with negative mass (relative to AdS 
with periodic identification) was found but it was 
conjectured to be the lowest energy solution
among all solutions with the same asymptotics.

Notice that the positivity of the gravitational energy implies
via the AdS/CFT  correspondence the 
positivity of the quantum QFT Hamiltonian at strong coupling.
This is a very strong conclusion since for a general 
AlAdS the dual QFT resides on a curved manifold, 
and in general even the very definition of a QFT on 
a curved manifold is subtle.  
Therefore, in general we expect that only a subclass of AlAdS 
spacetimes is subject to a positivity theorem.
One might in fact turn things around and view 
our discussions as giving a criterion for the selection 
of good boundary conditions. 

This paper is organized as follows. In the next section we discuss 
the definition of asymptotically locally AdS spacetimes and in 
section 3 the definition of energy for such spacetimes. 
The spinorial energy of Witten and Nester is reviewed in 
section 4. In section 5, we construct asymptotic 
solutions of the Witten equation and  use them 
in section 6 to compute a regulated version of the 
Witten-Nester energy for AlAdS spacetimes. This 
leads to a number of necessary conditions for the existence
of such an energy. In section 7 we compare the finite part 
of the Witten-Nester energy with the holographic energy.
In section 8 we specialize to AAdS spacetimes and in 
section 9 we illustrate subtleties related to some global issues 
by discussing two examples, the extremal BTZ black hole
and the AdS soliton. We conclude with a discussion of our results in 
section 10.
In order to keep the line of argument clear,
we have moved most of the technical details to a 
series of appendices.

\section{Asymptotically locally AdS spacetimes} \label{AlAdS}

We discuss in this section the definition of asymptotically locally
anti-de Sitter (AlAdS) spacetimes. More details can be 
found in \cite{Skenderis:2002wp,PS3} and the mathematics reviews
\cite{Graham,anderson}. In this paper,
we restrict our attention to the case of pure gravity
but the method can be generalized to include matter.

The most general asymptotic solution of Einstein's equations with
negative cosmological constant takes the form \cite{FG} 
\bea \label{coord} 
&&ds^2=G_{\m \n} dx^\m dx^\n = \frac{d z^2}{z^2} +
\frac{1}{z^2} g_{ij}(x,z) dx^i dx^j, \nonu &&g(x,z)=g_{(0)} + z
g_{(1)} \cdots + z^{d} g_{(d)} + h_{(d)} z^{d} \log z^2 + \cdots
\eea
In these coordinates $z=0$ is the location of the conformal
boundary of spacetime and $g_{(0)}$ is an arbitrary non-degenerate
metric (which represents the conformal structure of the boundary).
Einstein equations determine uniquely all coefficients in
(\ref{coord}) except for the transverse traceless part of $g_{(d)}$
\cite{FG,HS,dHSS} (see appendix A of \cite{dHSS} for concrete
expressions). A short computation reveals that 
the Riemann tensor of the metric (\ref{coord}) is asymptotically 
equal to 
\be \label{curAdS}
R_{\m \n \k \l} = %{1 \over l^2}
(G_{\m \l} G_{\n \k} - G_{\k \m} G_{\n \l})(1 + \co(z)) 
\ee 
where the cosmological constant is normalized as $\L =-d(d-1)/2$
(i.e. we set the AdS radius equal to one). Thus the
leading form of the Riemann tensor is exactly the same as the
Riemann tensor of the $AdS_{d+1}$ spacetime. We will call
solutions with this property ``asymptotically locally AdS''
(AlAdS) spacetimes. All solutions of pure gravity with negative
cosmological constant are of this form. Notice that we do not
require the conformal structure of (\ref{coord}) to be that of
$AdS_{d+1}$. Spacetimes with this conformal structure are called
``asymptotically AdS'' \cite{AM,Henneaux:1984xu}.

Recall that $AdS_{d+1}$ is conformally flat and this implies
\cite{Skenderis:1999nb} that $g_{(0)}$ is also conformally flat and
the expansion (\ref{coord}) terminates at order $z^4$,
\be \label{AdScoef}
g(x,z) = \left(1+\frac{z^2}{2} g_{(2)} g_{(0)}^{-1} \right) g_{(0)}
 \left(1+\frac{z^2}{2} g_{(0)}^{-1} g_{(2)}  \right)
\ee
with
\bea 
d=2: \qquad g_{(2)ij} &=& -\half(R g_{(0)ij} + t_{ij}), 
\qquad \nabla^i t_{ij}=0, \quad t^i_i=R, \label{AdS_d2} \\
d \neq 2: \qquad  g_{(2)ij} &=& -\frac{1}{d-2} (R_{ij} - \frac{1}{2 (d-1)} R
g_{(0)ij}), \nonumber
\eea
where $R_{ij}$ is the Ricci tensor of $g_{(0)}$ and the 
transverse traceless part of $t_{ij}$ 
is not determined by the asymptotic analysis.
$g_{(0)}$ may be chosen to
be the standard metric $\mathbf{R} \times S^{d-1}$. By definition, $AAdS_{d+1}$
spacetimes have the same boundary conformal structure as $AdS_{d+1}$.
This implies that all coefficients up to $g_{(d)}$
are the same as those for $AdS_{d+1}$, but $g_{(d)}$ is different.
For $AAdS_{d+1}$ spacetimes the logarithmic term in (\ref{coord}) is absent.

AlAdS spacetimes have an arbitrary conformal structure $[g_{(0)}]$
and a general $g_{(d)}$, the logarithmic term is in general
non-vanishing, and there is no {\it a priori} restriction on the
topology of the conformal boundary. The mathematical structure of
these spacetimes (or their Euclidean counterparts) is 
currently under investigation in the mathematics community, see
\cite{anderson} and references therein. For instance, it has not
yet been established how many, if any, global solutions exist
given a conformal structure, although given (sufficiently regular)
$g_{(0)}$ and $g_{(d)}$ a unique solution exists in a thickening
$B \times [0,\epsilon)$ of the boundary $B$. On the other hand,
interesting examples of such spacetimes have appeared in the
literature, see \cite{anderson} for a collection
of examples. One of the motivations for
the current work is to derive physically motivated conditions on
the possible conformal structures $[g_{(0)}]$.

A very useful reformulation of the asymptotic analysis can be
achieved by observing that for AlAdS spacetimes the radial
derivative is to leading order equal to the dilatation
operator \cite{PS1,PS2}.
That is to say, if we write the metric 
in the form
\be \label{gr_gauge}
ds^2 = dr^2 + \gamma_{ij}(x,r) dx^i dx^j
\ee
which is related to (\ref{coord}) by the coordinate transformation
$z=\exp(-r)$, then
\be
\partial_r = \delta_D + \co(e^{-r})
\ee
where $\d_D$ is the dilatation operator.
For pure gravity
\be \label{dil}
\d_D = \int d^d x  2 \g_{ij} \frac{\d}{\d \g_{ij}}.
\ee
When matter fields are present $\d_D$ contains additional terms
according to the Weyl transformation of the corresponding
boundary fields (see \cite{PS1,PS3} for examples).
The asymptotic analysis can now be very effectively
performed \cite{PS1} by expanding all objects in eigenfunctions of the
dilatation operator and organizing the terms in the field equations according
to their dilatation weight.

For the case of pure gravity, the main object is the extrinsic curvature
$K_{ij}$  of constant-$r$ slices. In the coordinates
where the metric is given by (\ref{gr_gauge}), the extrinsic curvature
is equal to
\be \label{defK}
K_{ij} = \half \dot{\g}_{ij},
\ee
where the dot indicates derivative w.r.t. $r$.
It admits the following expansion in terms of eigenfunctions
of the dilatation operator,
\be \label{K_exp}
K^{i}_{\,j}[\g] = \d^{i}_{\,j} +K_{(2)}{}^{i}_{j}
+K_{(4)}{}^{i}_{j}+ \cdots +
K_{(d)}{}^{i}_{j}
+\til{K}_{(d)}{}^{i}_{j} (-2r) + \cdots
\ee
where all terms but $K_{(d)}{}^i_j$ transform
homogeneously with the weight indicated by their subscript,
\be
\d_D K_{(n)}{}^i_j = - n  K_{(n)}{}^i_j, n<d, \quad
\d_D \til{K}_{(d)}{}^i_j = - d  \til{K}_{(d)}{}^i_j
\ee
and $K_{(d)}{}^i_j$ transforms anomalously,
\be
\d_D K_{(d)}{}^i_j = - d  K_{(d)}{}^i_j - 2 \til{K}_{(d)}{}^i_j.
\ee
Notice that $\til{K}_{(2k+1)}{}^{i}_{j}=0$.
The radial derivative admits a similar expansion:
\bea \label{exp_pa}
\pa_r &=& \d_D + \pa_{(2)}
%\sum_{k=1}^{[\frac{d+1}{2}]-1}\pa_{r(2k)} + \pa_{(d)}-(2 r) \tilde{\pa}_{(d)}
+ \cdots \nonumber \\
&=& \int d^d x \dot{\g}_{ij} \frac{\d}{\d \g_{ij}}
= 2 \int d^d x K_{ij} \frac{\d}{\d \g_{ij}} 
=\d_D + \int d^d x K_{(2) ij} \frac{\d}{\d \g_{ij}} +
%\nonumber \\
%&=& \d_D + \int d^d x \left( \sum_{k=1}^{[\frac{d+1}{2}]-1} K_{(2k) ij}
%+ K_{(d) ij} - (2 r) \til{K}_{(d) ij} \right) \frac{\d}{\d \g_{ij}} +
\cdots
\eea
where we used the chain rule and (\ref{defK})-(\ref{K_exp}).

Inserting these expansions in Einstein's equations
and grouping  terms with the same weight together leads to a
number of recursion relations that can be solved to uniquely determine
all coefficients except for the traceless divergenceless part of
$K_{(d)}{}^i_j$ \cite{PS1}.

The coefficients $K_{(n)ij}, \til{K}_{(d)ij}$ determine
the coefficients $g_{(n)}, h_{(d)ij}$ in (\ref{coord})
and vice versa. The precise relations have been worked out in \cite{PS1}
and we list them here for up to $n=4$,
\bea \label{k_g}
\g_{ij}[g_{(0)}] &=& \frac{1}{z^2} (g_{(0)ij} + g_{(2)ij} z^2 + \cdots
+ z^{d} g_{(d)ij} + h_{(d)ij} z^{d} \log z^2 + ...) \nonumber \\
K_{(2)ij}[g_{(0)}] &=& - g_{(2)ij} \\
K_{(3)ij}[g_{(0)}] &=& - \frac{3}{2}g_{(3)ij},  \qquad {\rm for} \ \ d=3 
\nonumber \\
K_{(4)ij}[g_{(0)}] &=& - 2 g_{(4)ij} - 3 h_{(4)ij} +
(g_{(2)}^2)_{ij} -\frac{1}{12} g_{(0)ij} (\Tr g^2_{(2)} - (\Tr g_{(2)})^2)
\nonumber \\
&&- \half g_{(2)ij} \Tr g_{(2)}, \qquad {\rm for} \ \ d=4
\nonumber \\
\til{K}_{(d)ij}[g_{(0)}] &=& - \frac{d}{2}  h_{(d)ij}. \nonumber
\eea
Explicit expressions for $g_{(n)}, h_{(n)}$ (for low enough $d$)
can be found in appendix A of \cite{dHSS} and expressions for $K_{ij}[\g]$
in \cite{PS1}.

Since the dilatation operator is  equal to the
radial derivative to leading order, the leading radial dependence
of a dilatation eigenfunction $f_{(k)}$ of weight $k$ is
equal to $\exp(-k r)$. It will
be useful to introduce the following ``hat'' notation for the leading
coefficient:
\be \label{hat}
f_{(k)} = e^{-k r} \hat{f}_{(k)}(x) (1 + \co(e^{-r}))
\ee
For instance, $\hat{\g}_{(-2)ij}(x)$ denotes the boundary metric
$g_{(0)ij}(x)$ and $\hat{K}_{(n)ij}=K_{(n)ij}[g_{(0)}]$.

\section{Energy of Asymptotically locally AdS spacetimes}

In gravitational theories energy is usually measured with respect to
a reference spacetime, but such a reference spacetime may 
not exist for general AlAdS spacetimes. 
In AlAdS spacetimes that possess an asymptotic
timelike Killing vector, however, one can do better: 
one can assign a mass in a way that is intrinsic to the spacetime,
as we review in this section.

We first note that all AlAdS
spacetimes possess a covariantly conserved energy momentum tensor constructed
from the metric coefficients (in general there are 
contributions from matter \cite{dHSS,howtogo,holren,Hollands:2005wt,PS3}, 
but we only discuss the pure gravity case in this paper),
\be \label{tij}
T_{ij} = -\frac{1}{\k^2} (K_{(d)ij} - K_{(d)} \g_{ij})
\ee
where $K_{(d)} = K_{(d)}{}^i_i$ and $\k^2 = 8 \pi G$. 
This energy momentum tensor
is equal to the variation of the gravitational on-shell
action supplemented by appropriate boundary counterterms
w.r.t. the boundary metric \cite{BK,dHSS,PS1}.
One can also derive (\ref{tij}) as a
Noether current associated with asymptotic (global) symmetries
of the bulk spacetime \cite{PS3}. When the bulk equations
of motion hold, it satisfies,
\be
\nabla^i T_{ij} =0, \qquad T^i_i = A,
\ee
where $A$ is the holographic anomaly ($A$ is non-vanishing 
only for even $d$ for the pure gravity case 
but when matter is present there may 
be additional conformal anomalies for all $d$ \cite{PS}).

Let us consider an AlAdS spacetime that possesses a vector that  
asymptotically approaches a  
conformal Killing vectors $\xi^i$ of the boundary metric
$g_{(0)}$ (see appendix B of \cite{PS3} for the precise fall off conditions). 
Conserved charges are now obtained
as,
\be \label{charges}
Q_h = -\int_{C_t \cap \pa M} dS_{i}\,T^{i}_{\;\;j}\,\x^{j}
\ee
where $C_t$ is an initial value hypersurface of the bulk
manifold. If the anomaly vanishes one can construct conserved
charges for all conformal Killing vectors of the boundary
metric. In particular, the energy is associated with a timelike 
Killing vector.

One can compute the value of the energy with following steps
(see also section 6 of \cite{PS3}):
\begin{enumerate}
\item Bring the bulk metric to the form (\ref{coord}) by changing
coordinates near $z=0$, and read off the coefficients $g_{(n)}$. From 
these coefficients, compute the $K_{(d)ij}$ coefficients
using (\ref{k_g}).
\item Compute the stress energy tensor $T_{ij}$ by substituting
$K_{(d)ij}$ in (\ref{tij}).
\item Plug in  $T_{ij}$ and the timelike Killing
vector $\xi^i$ of $g_{(0)}$ in (\ref{charges}).
\end{enumerate}

Let us illustrate this procedure by computing
the mass of $AdS_5$.
We already reported the result for step 1 in (\ref{AdScoef}).
Substituting in (\ref{tij}) we obtain the stress energy tensor 
\cite{Skenderis:2000in}
\be
T_{ij} = \frac{1}{64 \p G_N} (4 \delta_{i,0}\delta_{j,0} + g_{(0)ij})
\ee
The boundary metric is in this case the standard metric on $R \times S^3$,
so the timelike Killing vector is $\xi=\pa/\pa t$.
Substituting in (\ref{charges}) we get
\be
M_{AdS_5}=\int d^3x \sqrt{g}  T_{00} = \frac{3 \pi}{32 G_N}.
\ee
In previous approaches \cite{Abbott:1981ff,AM,Henneaux:1984xu,Hawking:1995fd}
one could only measure the energy of
spacetimes {\it relative} to $AdS_5$. Here we see that
we can compute the mass for $AdS_5$ itself. The fact that the mass is 
non-zero is due to the presence of the conformal anomaly
(which is related to IR divergences of the on-shell action).
Its value is {\it exactly equal} to the Casimir
energy of $N=4$ SYM  on $S^3$ \cite{BK}.

The purpose of this work is to analyze under which conditions
the energy defined holographically is bounded from below. To
answer this questions we will connect the holographic energy
to the spinorial energy of Witten and Nester that is manifestly
positive definite.

\section{Positivity of energy}

Witten's positive energy theorem \cite{Witten:1981mf}
is motivated by the fact that in supersymmetric theories
the Hamiltonian is the square of supercharges. This
implies that there is an expression for the energy in terms
of spinors and that the energy is positive definite.
The construction below imitates the
supersymmetric argument but does not require supersymmetry.

Given an antisymmetric tensor $E^{\m \n}$, one can always obtain
an identically covariantly conserved current (i.e. the conservation
does not require use of field equations)
\be
j^{\m}={\cal D}_{\n} E^{\n\m}  \qquad \Rightarrow \qquad
{\cal D}_\mu j^\mu = -2 R_{\m \n} E^{\m \n} =0,
\ee
where ${\cal D}_\mu$ is the covariant derivative associated
with the bulk metric $G$.
Integrating the time component of this current
 over a spacelike hypersurface $C_t$, we obtain a conserved charge
\be  \label{cons1}
Q = \int_{C_t} d\S_\mu j^\mu = \int_{C_t} d^{d}x
 \,\sqrt{\tG}
\; n_{\m} {\cal D}_{\n}E^{\n\m} \;,
\ee
where $\tG$ is the induced metric on the hypersurface $C_t$ and
$n_\mu$ is the unit normal of $C_t$. Using
Stokes' theorem\footnote{Notice that
$n_{\m} {\cal D}_{\n}E^{\n\m}={}^t \mathcal{D}_\n (n_{\m} E^{\n\m})$,
where ${}^t \mathcal{D}$ is the covariant derivative on $C_t$.}
and assuming that the spacetime has a single boundary, we
obtain a formula for the charges as an integral at infinity
\be \label{cons2}
Q = \int_{C_t \cap \pa M} d \S_{\m \n} E^{\n \m}
= \int_{C_t \cap \pa M}  d^{d{-}1}\!x \sqrt{{}^tg_{(0)}}\; n_\m l_\n E^{\n \m},
\ee
where ${}^tg_{(0)}$ is the induced metric on $C_t \cap \pa M$ and
$l^\m$ is the outward pointing unit normal of the boundary $\pa M$.

The Witten-Nester spinorial energy $E_{WN}$ \cite{Witten:1981mf,Nester}
is derived using the following
 antisymmetric tensor constructed from a spinor fields
\(\e\),
\be
E^{\m\n}=\frac{1}{\k^2} (\bare \G^{\m\n\r}\na_{\r}\e + c.c.)
%&=&\bare_{1}\G^{\m\n\r}\na_{\r}\e_{2}
%-\overline{\na_{\r}\e_{1}}\G^{\m\n\r}\e_{2},
\ee
where
\be
\nabla_\mu = {\cal D}_\mu + \half \G_\mu,
\ee
is the AdS covariant derivative (as noted before,
we set the AdS scale $l=1$ throughout this paper).
A standard computation (see, for instance, \cite{GHW}
for details) that uses the bulk equations of motion\footnote{As mentioned 
earlier, we consider the case of pure
gravity in this paper. The positivity of the spinorial energy
continues to hold for gravity coupled to matter with a stress energy tensor
that satisfies the dominant energy condition.}
\be \label{feq}
R_{\m \n} - \half (R-2 \L) G_{\m \n}=0,
\ee
yields
\be \label{bulkintegrand}
n_{\m}{\cal D}_{\n}E^{\n\m} = 2 \left({(\na_{\hat{\a}}\e)}^{\dag}
\eta^{\hat{\a}\hat{\b}} (\na_{\hat{\b}}\e) -
(\G^{\hat{\a}} \nabla_{\hat{\a}} \e)^\dagger
(\G^{\hat{\b}} \nabla_{\hat{\b}} \e) \right),
\ee
where the indices $\a, \b$ run through all values except time and the hat
indicates a flat index, e.g. $\nabla_{\hat{\a}}
= E_{\hat{\a}}^\m \nabla_\m$ with $E_{\hat{\a}}^\m$ being the inverse
vielbein, see appendix \ref{conventions} for our conventions.
It follows that if there exists a regular spinor $\e$ on $C_t$
satisfying the Witten equation,
\be
\G^{\hat{\a}} \nabla_{\hat{\a}} \e =0,
\ee
the Witten-Nester energy is positive definite,
\be
E_{WN} \geq 0.
\ee
Furthermore, the equality holds iff the Witten spinor is covariantly constant
w.r.t. to the AdS connection,
\be
E_{WN}=0 \qquad \Leftrightarrow \qquad \nabla_{\hat{\a}} \e =0.
\ee

On the other hand, the value of $E_{WN}$ depends only on the
asymptotics of the Witten spinor as follows from (\ref{cons2}).
We would like to compute this energy for general AlAdS spacetime.
To regulate potential IR divergences we introduce a regulating
surface $\S_r$. The regulated energy is now given by
\be \label{E_reg0}
E_{WN}[r]= \int_{C_t \cap \S_r}
d^{d{-}1}\!x \sqrt{{}^t\g}\; n_\m l_\n E^{\n \m}
=-\frac{1}{\k^2} \int_{C_t \cap \S_r} d^{d{-}1}\!x \sqrt{{}^t\g}\
(\e^\dagger \G^{\hatr} \G^{\hata} \nabla_{\hat{a}}
\e + c.c.)
\ee
where we used that in our case
$n_\mu = E_\mu^{\hat{t}}$ and $l_\mu = E_\mu^{\hat{r}}$,
see appendix \ref{slices}. To compute this expression
we need to know asymptotic solutions of the Witten equation.

\section{Asymptotic solutions of the Witten equation}

We would like to obtain asymptotic solutions of the
Witten equation,
\be \label{weq}
\sla \nabla \e \equiv (\G^{\hatrr}\na_{\hatrr}+\G^{\hata}\na_{\hata})\e=0.
\ee
This is obtained by expanding all quantities
in terms of dilatation eigenfunctions, as in the asymptotic
analysis of the bulk equations of motion \cite{PS1} reviewed
in section 2. We present the details in appendix \ref{as_exp}.
In particular, we find the Witten operator
$\sla \nabla$ admits the following expansion,
\be
\sla \nabla = \sd_{(0)} + \sd_{(1)}
+ \sum_{k=1}^{[\frac{d-1}{2}]} \sd_{(2k)} +
\sd_{(d)} + (-2r)\tilde{\sd}_{(d)} + \cdots,
\ee
where the explicit expressions can be found in appendix \ref{as_exp}
($[k]$ denotes the integer part of $k$).
We only quote here the first two terms
\be
\sd_{(0)} =( \dil\,+\frac{d-1}{2})\Gr+\frac{d}{2}, \qquad
\sd_{(1)} =\G^{\hata}D_{\hata}
\ee
and note that $\tilde{\sd}_{(d)}$ is zero when $d$ is odd.

Let us now consider a spinor with the asymptotic expansion
\be \label{sp_exp}
\e=\e_{(m)}+\e_{(m+1)}+\e_{(m+2)}+ \cdots +\e_{(m+d)}
+(-2r)\,\tilde{\e}_{(m+d)}+ \cdots \;,
\ee
where the coefficients transform as their subscript indicates,
\be
\d_D \e_{(n)} = - n \e_{(n)}, \qquad \d_D \til{\e}_{(m+d)} = - (m+d)
\til{\e}_{(m+d)}
\ee
except for $\e_{(m+d)}$ which transforms anomalously,
\be
\d_D \e_{(m+d)} = - (m+d) \e_{(m+d)} - 2 \til{\e}_{(m+d)}.
\ee

Inserting (\ref{sp_exp}) in the Witten equation and collecting
terms of the same
weight, we get a series of equations. The equation for the lowest
order component reads
\be
\sd_{(0)}\, \e_{(m)} = 0.
\ee
This implies that either
\be \label{be1}
 m= \mhalf, \qquad \e_{(-\frac{1}{2})}= \Pm \e_{(-\frac{1}{2})}
\ee
or
\be \label{be2}
 m= d \mhalf, \qquad \e_{(d-\frac{1}{2})}= \Pp \e_{(d-\frac{1}{2})},
\ee
where \(P^{\pm}=\frac{1}{2} (1\pm\Gr)\) are projection operators. 
The Witten spinors
with leading behavior as in (\ref{be2}) fall off too fast at
infinity to contribute to $E_{WN}$, and therefore we 
consider only the solution with leading behavior as in (\ref{be1}) from now on.
Notice, however, that a Witten spinor which is regular in the interior
may require a linear combination of the two asymptotic solutions.

The remaining equations read
\bea \label{generalsol1}
\sd_{(0)}\,\e_{(-\half+k)}&=&
-\left(\sd_{(1)}\,\e_{(k-\frac{3}{2})}+\sum_{l=1}^{[\frac{k}{2}]}
\sd_{(2l)}\e_{(-\half+k-2l)}\right), 
\qquad k=1, 2, \ldots, d{-}1 \label{eq_k} \\
\sd_{(0)}\,\til{\e}_{(-\half+d)}&=&-\til{\sd}_{(d)}\e_{(-\half)}\;,
\label{eq_til} \\
\sd_{(0)}\, \e_{(-\half+d)} &=& -\left(\sd_{(1)} \e_{(d-\frac{3}{2})}
+\sum_{l=1}^{[\frac{d-1}{2}]}
\sd_{(2l)}\e_{(-\half+d-2l)} + \sd_{(d)} \e_{(-\half)}\right). \label{eq_d}
\eea
Using the commutation relations between $\sd_{(n)}$ and $P^{\pm}$
listed in (\ref{begin}) we conclude
 \be\label{eind}
\Gr \e_{(\mhalf+n)} = (-1)^{n+1}\, \e_{(\mhalf+n)}
\;\;,\;0\leq n<d\,.\ee

Equations (\ref{generalsol1}) can be solved iteratively  to determine
locally all coefficients in terms of $\e_{(-\half)}$ provided $\sd_{(0)}$
is invertible. The zero modes of $\sd_{(0)}$ are given in
(\ref{be1}) and (\ref{be2}), so starting from $\e_{(-\half)}$
one can determine all coefficients except for $P^+ \e_{(-\half + d)}$
which is left undetermined. The result is\footnote{
Use $
\left(( \dil\,+\frac{d-1}{2})\Gr-\frac{d}{2} \right) \sd_{(0)}
\e_{(-\half +k)} = k (d-k) \e_{(-\half +k)}$.}
\bea
&&\e_{(-\half+k)} = -c(k)
\left(\sd_{(1)}\,\e_{(k-\frac{3}{2})}+\sum_{l=1}^{[\frac{k}{2}]}
\sd_{(2l)}\e_{(-\half+k-2l)}\right), \\
&&{\rm with}
\quad c(2l)=\frac{1}{2l}, \quad c(2l{+}1){=}\frac{1}{d-(2l+1)},
\qquad k=1, 2, \ldots, d{-}1. \nonumber
\eea
Later on we will need the explicit form for $k=1$:
\be \label{sol1}
\e_{(\frac{1}{2})}=-\frac{1}{d-1}\,\G^{\hata}D_{\hata}\,\e_{(\mhalf)},
\ee

The solution of (\ref{eq_til}) and (\ref{eq_d}) depends on
whether $d$ is even or odd:\newline
$\bullet\ d$ even
\bea
\Pm\,\til{\e}_{(\mhalf+d)}
&= & -\frac{1}{d} \til{\sd}_{(d)}\e_{(\mhalf)} \nonumber \\
\Pp\,\til{\e}_{(\mhalf+d)} &=&0 \\ \nonumber
\Pm\,\e_{(\mhalf+d)}
&=& - \frac{1}{d} \left( \G^{\hata}D_{\hata} \e_{(d-\frac{3}{2})}
+2\til{\e}_{(\mhalf+d)}+
\sum_{k=1}^{d/2} \sd_{(2k)} \e_{(\mhalf+d-2k)} \right) \\
\Pp \e_{(\mhalf+d)} &&\;\; {\rm undetermined} \nonumber
\eea
$\bullet\ d$ odd
\bea
\Pm\til{\e}_{(\mhalf+d)} &=& 0 \nonumber \\
\Pp \til{\e}_{(\mhalf+d)}  &=& \mhalf \left(
\sd_{(1)} \e_{(d-\frac{3}{2})}+
\sum_{k=1}^{(d-1)/2}\sd_{(2k)} \e_{(\mhalf+d-2k)} \right) \nonumber\\
\Pm \e_{(\mhalf+d)}&=& -\frac{1}{d}\,\sd_{(d)} \e_{(\mhalf)} \\
\Pp \e_{(\mhalf+d)} &&\;\; {\rm undetermined}. \nonumber
\eea
Having determined the most general asymptotic solution of 
the Witten equation,
we next turn to the computation of the Witten-Nester energy.

\section{Witten-Nester energy}

We are now in the position to compute the Witten-Nester energy.
Recall that the regulated expression is given by
\be \label{E_reg}
E_{WN}[r]=-\frac{1}{\k^2} \int_{C_t \cap \S_r} d^{d{-}1}\!x \sqrt{{}^t\g}\
(\e^\dagger \G^{\hatr} \G^{\hata} \nabla_{\hat{a}}
\e + c.c.)
\ee
where $r$, the position of the radial slice, is the regulator.
Using the asymptotic expansion derived in the previous section
we obtain
\bea
q &\equiv& - \e^{\dag} \Gr \G^{\hata} \na_{\hata} \e \nonumber \\
&=&q_{(-1)}+q_{(0)}
+q_{(1)}+ \cdots +q_{(d-1)}
+(-2r)\til{q}_{(d-1)}+ \cdots \;\;. \label{decompose}
\eea
All terms up to $q_{(d-1)}$ give divergent contributions in (\ref{E_reg})
as $r \to \infty$.
Therefore, for $E_{WN}$ to be
well-defined, these terms should vanish. Similar divergences
were found in the on-shell action in \cite{HS} and there
they were canceled by means of boundary counterterms.
In the present context, however, we want to maintain the
manifest positivity of $E_{WN}$ so instead of
adding counterterms we view the vanishing of the divergent
terms as conditions imposed on the asymptotic data.
In other words, our results show that only for a subset of AlAdS spacetimes
the Witten-Nester energy is well-defined.
We should add here that our discussions
do not exclude the possibility that a modified Witten-Nester energy
exists that is manifestly positive and is well defined
for a wider class of AlAdS spacetimes.

The explicit form of $q_{(n)}$ is
most easily obtained by using (\ref{d_a_begin}).
Using the alternating chirality of the spinors $\e_{(k)}$
(\ref{eind}) we conclude
\be
q_{(-1)} = q_{(2l)}
=0 \;\;{\rm for}\;\,l= 0, 1, \ldots \quad  2l\neq d-1\,,
\ee
The odd powers however are generically non-zero,
\bea\nonumber
q_{(2n-1)} &= &-(d-1)\;\sum_{k=0}^{n-1}¥\,
\dage_{(2n-2k-\frac{3}{2})}\e_{(\frac{1}{2}+2k)}
+\sum_{k=0}^{2n-1}\;(-1)^{k+1}\,\dage_{(2n-k-\frac{3}{2})}
\G^{\hata}D_{\hata} \e_{(\mhalf+k)¥}\\ \label{2n-1}
&&-\frac{1}{2} \sum_{k=1}^{n}\;\sum_{l=0}^{2n-2k}\;K_{(2k)}{}^{\hatj}_{\;\hata}
\;\dage_{(2n-2k-l\mhalf)} \G^{\hata}\G_{\hatj}\,
\e_{(l\mhalf)}
\eea
for \(n = 1, 2, \ldots , [\frac{d-1}{2}]\).

The result for the terms of order $(d-1)$ depends on whether $d$ is
even or odd, \newline
$\bullet\ d$ odd
\bea \label{d-1}
q_{(d-1)} &= &
\mhalf K_{(d)}{}^{\hatj}_{\;\hata}
\dage_{(\mhalf)}\G^{\hata}\G_{\hatj}\e_{(\mhalf)}\,\\
\til{q}_{(d-1)}&=& 0 \nonumber \;,
\eea
$\bullet\ d$ even
\bea  \label{d-1*}
q_{(d-1)} &= & \mhalf K_{(d)}{}^{\hatj}_{\;\hata}
\dage_{(\mhalf)}\G^{\hata}\G_{\hatj}\e_{(\mhalf)}
+ \half A_{(d-1)} \\
\til{q}_{(d-1)}&=& \mhalf \til{K}^{\hatj}_{\;\hata(d)}
\dage_{(\mhalf)}\G^{\hata}\G_{\hatj}\e_{(\mhalf)} \nonumber
\eea
where we separated out in $q_{(d-1)}$ the term that depends
on the coefficient $K_{(d)ij}$ which is not determined by the asymptotic
analysis. The remaining terms are given by
\bea
\half A_{(d-1)} &=& -(d-1)\;\sum_{k=0}^{d/2-1}¥\,
\dage_{(d-2k-\frac{3}{2})}\e_{(\frac{1}{2}+2k)}
+\sum_{k=0}^{d-1}\;(-1)^{k+1}\,\dage_{(d-k-\frac{3}{2})}
\G^{\hata}D_{\hata} \e_{(\mhalf+k)¥} \nonumber \\
&&-\frac{1}{2} \sum_{k=1}^{d/2-1}\;
\sum_{l=0}^{d-2k}\;K_{(2k)}{}^{\hatj}_{\;\hata}
\;\dage_{(d-2k-l\mhalf)} \G^{\hata}\G_{\hatj}\,
\e_{(l\mhalf)} \label{ad}
\eea

To summarize, we have the following result
\bea
d\ {\rm odd}: && q = \sum_{l=1}^{\frac{d-1}{2}}q_{(2l-1)}
+ q_{(d-1)} + \cdots \\
d\ {\rm even}: &&  q = \sum_{l=1}^{\frac{d}{2}-1}q_{(2l-1)}
+ (-2r)\,\til{q}_{(d-1)} + q_{(d-1)} +\cdots
\eea
where the various coefficients are given in
(\ref{2n-1}), (\ref{d-1}) and (\ref{d-1*}).

In order for the Witten energy to be well defined we need the integral
of the divergent coefficients be zero.
Recall that $\tilde{K}_{(d)ij}$ is the metric variation of the
conformal anomaly \cite{dHSS} and vanishes when the
boundary metric is conformally Einstein \cite{FG}, i.e. when there
exists a representative of the boundary conformal structure
$g_{(0)}$ that satisfies Einstein's equations (with or without
cosmological constant). So we conclude that a sufficient condition for the vanishing of the
``logarithmic'' divergence\footnote{Recall that $-2r=\log z^2$
and $\exp(k r)=z^{-k}$
so $(-2 r) \til{q}_{d-1}$ and $q_{(2 n-1)}$ are analogous to the logarithmic
and power-law divergences
in the on-shell action.} (which is present only in
even dimensions) is that the boundary metric is
conformally Einstein.

The ``power-law'' divergences $q_{(2n-1)}$ impose further conditions
on the asymptotic data, namely the boundary geometry
should be such that spinors $\e_{(-\half)}$ satisfying
specific differential equations exist. In $d=2$ there is no such divergence.
For  $d=3, 4$ the only divergent term is $q_{(1)}$. This results in the
following condition\footnote{$q_{(1)}$ is equal to $-\e_{(-\half)}^\dagger$
times the l.h.s. of (\ref{pl_cond}).}
\be \label{pl_cond}
\left(\frac{1}{(d-1)} (\G^{\hata} D_{\hata})^2
+ \half K_{(2)}{}^{\hatj}_{\hata}
\G^{\hat{a}} \G_{\hatj}\right)
\e_{(-\half)} =0
\ee
where
\be \label{k2}
K_{(2)ij}= \frac{1}{(d-2)}\left(R_{ij} -\frac{1}{2(d-1)} R \g_{ij}\right).
\ee
This condition is not Weyl covariant
but one can understand this as a consequence  of the invariance of the
Witten-Nester energy under diffeomorphisms, as we discuss
in appendix \ref{app_weyl}. We are not aware of a classification of
manifolds that admit such spinors, but we will discuss examples
below where this condition is satisfied. The conditions $q_{(2n-1)}$
for $n>1$ will only be discussed for AAdS spacetimes.

\section{Holographic energy vs Witten-Nester energy \label{Hol_WN}}

In the previous section, we discussed necessary conditions
for the Witten-Nester energy to be well defined. We assume now that
these conditions hold and we discuss how the finite part compares
with the holographic energy. 

Using (\ref{E_reg})-(\ref{decompose})-(\ref{d-1})-(\ref{d-1*}), we get
\be
E_{WN}= \label{E_fin}
\frac{1}{2 \k^2} \int_{C_t \cap \S_r} d^{d{-}1}\!x \sqrt{{}^t\g}\
K_{(d)}{}^{\hatj}_{\;\hata}
\dage_{(\mhalf)}\G^{\hata}\G_{\hatj}\e_{(\mhalf)}
-\frac{1}{2 \k^2} \int_{C_t \cap \S_r} d^{d{-}1}\!x \sqrt{{}^t\g}\ A_{(d-1)}
+c.c.
\ee
where $A_{(d-1)}$ is non-zero only for even $d$. A simple algebra shows that 
\be
K_{(d)}{}^{\hatj}_{\;\hata}
\dage_{(\mhalf)}\G^{\hata}\G_{\hatj}\e_{(\mhalf)}
= \k^2 T^{\hatt}{}_{\hati} \bar{\e}_{(-\half)} \G^{\hati} \e_{(-\half)},
\ee
where $T_{ij}$ is the holographic stress energy tensor (\ref{tij}).
It follows 
\bea
&&d\ {\rm odd:} \qquad \qquad E_{WN} = E_h,  \\
&&d\ {\rm even:} \qquad \qquad E_{WN} = E_h  - E_0, 
\eea
provided $\hat{\e}_{(-\half)}$ 
is chosen such that
\be \label{kil_vec}
\xi^{i}= \bar{\hat{\e}}_{(-\half)} \G^{i} \hat{\e}_{(-\half)}
\ee
is a timelike Killing vector of the boundary metric $g_{(0)}$.
(The hat notation explained in (\ref{hat}))
Notice that $g_{(0)}$ must have a timelike Killing vector
in order to define energy.
We show in appendix \ref{Kil_Wit} that if the Witten spinor 
is asymptotically a Killing spinor then (\ref{kil_vec}) is 
automatically a timelike or null 
conformal Killing vector of the boundary metric.
In the more general case we discuss here Killing spinors may 
not exists even asymptotically, but $g_{(0)}$ can have  
a timelike Killing vector. 
In this case (\ref{kil_vec}) is viewed
as an additional condition on $\hat{\e}_{(-\half)}$.

The additional term for even $d$, i.e. for odd dimensional bulk
spacetimes, is equal to 
\be \label{eo_a}
E_0 = - \frac{1}{\k^2} \int_{C_t \cap \pa M} {\rm Re} (\hat{A}_{(d-1)}),
\ee
where $\hat{A}_{(d-1)}$ is given in (\ref{ad}).
It depends only on asymptotic data and 
is a bounded quantity.  For general 
$AlAdS_3$ and $AlAdS_5$ spacetimes $E_0$ is given by
\bea
&&\bullet d=2 \qquad \qquad \half \hat{A}_{(1)} = |\pa_\f 
\hat{\e}_{(-\half)}|^2 \\
&&\bullet d=4 \qquad \qquad \half\hat{A}_{(3)} = -\frac{1}{12} 
|\hat{K}^{\hatj}_{\;\;\hata(2)} \G^{\hata}\G_{\hatj} \hat{\e}_{(\mhalf)} |^2 
\label{a4}
\eea
where in deriving (\ref{a4}) we used the finiteness condition (\ref{pl_cond}),
$\f$ stands for the spatial boundary coordinate of $AlAdS_3$
and $K_{(2)ij}$ is given in (\ref{k2}).
In the next section we will derive $E_0$ for $AAdS_{d+1}$ spacetimes.

This leads us to the main result of this paper. 
Consider $AlAdS$ spacetimes where in addition to the boundary conformal
structure\footnote{
As discussed in detail in \cite{PS3} when the conformal anomaly 
does not vanish {\em identically} one needs to pick 
a specific representative $g_{(0)}$ in order to define the theory.} 
$[g_{(0)}]$ we also specify a boundary spinor $\hat{\e}_{(-\half)}(x)$.
We require that $(g_{(0)}, \hat{\e}_{(-\half)})$ 
are such that 
(i) the no-divergence conditions derived in the 
previous section are satisfied, 
(ii) $\xi^i$ in (\ref{kil_vec}) is a timelike Killing vector of $g_{(0)}$,
(iii) a regular Witten spinor approaching $\e_{(-\half)}$
asymptotically exists and 
(iv) the bulk spacetime  
has a single boundary or if there are more than one boundary
the other boundaries should give vanishing contribution to $E_{WN}$.

The holographic energy of $AlAdS$ spacetimes with such 
an asymptotic structure
is bounded from below
\bea
AlAdS_{2k}: \qquad &&E_h \geq 0 \label{ads_even} \\
AlAdS_{2k+1}: \qquad &&E_h \geq E_0 
\label{ads_odd}
\eea
Spacetimes saturating the bound may be considered as ``the ground 
state'' among all spacetimes with the same asymptotic data.

Notice that $E_0$ depends on $\hat{\e}_{(-\half)}$ so if 
$\hat{\e}_{(-\half)}$ is not fixed uniquely by our requirements,
$E_0$ in (\ref{ads_odd}) should be understood to be the maximum
among all choices. The fact that the bound in odd dimensions 
is non-zero 
is related to the fact that the Witten-Nester energy vanishes 
for supersymmetric solutions, but the holographic energy 
may not be zero, essentially because of the presence of the 
conformal anomaly. In fact $E_0$ 
for AAdS spacetimes is related via AdS/CFT to the Casimir energy of the 
dual QFT.  We discuss this further in the next section.

We finish this section with a few remarks. If the boundary 
metric has additional (conformal) isometries the Witten-Nester construction 
can be generalized to include all conserved charges. This is discussed
for AAdS spacetimes in \cite{GHW} (see also the recent discussion 
in \cite{Freedman:2003ax}). In such cases 
we expect exact agreement between the 
Witten-Nester charges and the holographic charges. We also expect that 
one is able to relax the last requirement, namely that all
contributions to $E_{WN}$ come from a single boundary.
The case of spacetimes
with horizons is discussed in \cite{GHHP}. Thus the main two 
requirements on the asymptotic structure are the no-divergence 
conditions and the global existence of Witten spinors.

\section{AAdS spacetimes}

In this section we restrict our attention to $AAdS_{d+1}$ spacetimes.
This case has been discussed previously in
\cite{GHW,GHHP,Davis:1986da,Hollands:2005wt}.
These spacetimes possess asymptotic
Killing spinors and we take the the Witten spinor
to approach such a spinor,
\be
\e_{W}(x,r) = \e_K(x,r)(1+\co(e^{-d r})),
\ee
where $e_K$ is the AdS Killing spinor given in (\ref{killing}).
Properties of AdS Killing spinors
are discussed in appendix \ref{AdSK}.

Recall that the asymptotics of $AAdS$ start differing from $AdS$ at
the normalizable mode order and the Witten-Nester energy is zero for $AdS_{d+1}$.
It follows that all divergent terms in $E_{WN}[AAdS]$ are zero. We will shortly demonstrate this
for up to $d=6$. Furthermore, since $E_0$ depends only on boundary data,
it is universal among all solutions with the same asymptotics. So to evaluate
it, it is sufficient to consider the case of $AdS_{d+1}$. We thus obtain
(using $E_{WN}[AdS]=0$)
\be
E_0 = E_h[AdS].
\ee
The energy of $AdS_{2p+1}$ (with boundary $R \times S^{2p-1}$)
can be evaluated using the results in appendix \ref{proof}
for any $p$ (for even dimensions $E_h[AdS_{2p}]=0$).
For up to $d=6$ one can actually compute the energy of $AdS_{d+1}$ with boundary metric
any conformally flat metric $g_{(0)}$ using the following formulae 
derived in \cite{dHSS} (the formula for $d=6$ corrects typos in
(3.21) of \cite{dHSS}),
\bea
d=2: && T_{ij} = \frac{1}{\k^2} [g_{(2)} - g_{(0)} \Tr g_{(2)}]_{ij} \\
d=4: && T_{ij} = \frac{1}{2 \k^2} [-g_{(2)}^2  
+ g_{(2)} \Tr g_{(2)}
-\frac{1}{2} g_{(0)} ((\Tr g_{(2)})^2 - \Tr g_{(2)}^2 )]_{ij} \nonumber \\
d=6: && T_{ij} = \frac{1}{4 \k^2} [g_{(2)}^3 - g_{(2)}^2 
\Tr g_{(2)} + \half g_{(2)} \left((\Tr g_{(2)})^2-\Tr g_{(2)}^2\right) \nonumber \\
&& \qquad \qquad 
+g_{(0)}\left(\half \Tr g_{(2)} \Tr g_{(2)}^2 - \frac{1}{3} \Tr g_{(2)}^3 
-\frac{1}{6} (\Tr g_{(2)})^3\right)]_{ij} \nonumber
\eea
Specializing these results to $g_{(0)}$ being the metric on $R \times 
S^{2 p-1}$ or using the results from
appendix \ref{proof} one obtains,
\be
E_0(d=2)=-\frac{\p}{\k^2}, \qquad E_0(d=4)=\frac{3 \p^2}{4 \k^2}, 
\qquad E_0(d=6)=-\frac{5 \p^3}{16 \k^2}
\ee
In the previous section we provided a formula for $E_0$  in terms of 
$A_{(d-1)}$, see (\ref{eo_a}).
It is a nice check on our computations that both computations give the 
same answer and we 
demonstrate this for up to $d=6$.

To explicitly check the cancellation of divergences and compute $E_0$ we 
need to know the coefficients $K_{(2n)ij}$
and $\e_{(m)}$. These are computed in appendix \ref{proof} and we give 
here only the relevant results
for the computation up to $d=6$,
\bea
K_{(2n)at}[\g]=0, \qquad \tilde{K}_{ij}[\g]=0, \qquad K_{ab}[\g]=\g_{ab} 
\sum_{n=0} \hat{k}_{(2n)} \g^{-n}, \nonumber \\
\hat{k}_{(0)}=1, \qquad \hat{k}_{(2)}=\frac{1}{2}, \qquad \hat{k}_{(4)} = 
- \frac{1}{8}, \qquad
\hat{k}_{(6)}=\frac{1}{16}, \label{k_coeff}
\eea
where
\be
\g_{ab}=\g g_{(0)ab}, \qquad \g = e^{2 r} \left(1-\frac{e^{-2 r}}{4}\right)^2,
\ee
and $g_{(0)ab}$ is the standard metric of $S^{2p-1}$.
For the expansion of the Killing spinor we get,
\bea
&&e_K(x,r)=\sum_{m=0} \hat{\e}_{(-\half+m)} \g^{-\frac{1}{2}(-\half+m)}, 
\nonumber \\
&&\hat{\e}_{(\frac{3}{2})}=\frac{1}{8}\hat{\e}_{(-\frac{1}{2})}, \qquad
\hat{\e}_{(\frac{5}{2})}=-\frac{1}{8}\hat{\e}_{(\frac{1}{2})}, \qquad
\hat{\e}_{(\frac{7}{2})}=-\frac{5}{128}\hat{\e}_{(-\frac{1}{2})}, \qquad
\hat{\e}_{(\frac{9}{2})}=\frac{7}{128}\hat{\e}_{(\frac{1}{2})}, \label{e_coeff}
\eea
where $\hat{\e}_{(\pm \frac{1}{2})}$ are
given in (\ref{eta}). Using (\ref{new_nablai}) one easily obtains that they 
satisfy,
\be
(\G^{\hata} \hat{D}_{\hata})^2 \hat{\e}_{(-\half)} =
-\frac{1}{4} (d-1)^2 \hat{\e}_{(-\half)}, \qquad
\hat{\e}_{(\half)}^\dagger \hat{\e}_{(\half)}
=\frac{1}{4} \hat{\e}_{(-\half)}^\dagger
\hat{\e}_{(-\half)} +\ {\rm total\ derivative}
\ee
and we normalize as $\hat{\e}_{(-\half)}^\dagger \hat{\e}_{(-\half)}=1$.

Using these results one can explicitly evaluate $q_{(1)}$ and $q_{(3)}$ and 
find that they
are equal to zero. Furthermore, $\tilde{q}_{(d-1)}=0$ for AAdS since the 
boundary metric
is conformally flat. This explicitly demonstrates that the Witten-Nester energy is well-defined
for up to $d=6$. Furthermore, one can also easily evaluate $A_{(d-1)}$ with result,
\be \label{a_k}
A_{(d-1)} = (d-1) \hat{k}_{(d)}.
\ee
This implies that
\be
E_{(0)}=E_h[AdS]
\ee
since the right hand side of (\ref{a_k}) is equal to $\k^2 T^{\hatt}_{\hatj} 
\xi^{\hatj}$,
where $T^i_j$ is the holographic stress energy tensor for $AdS_{2p+1}$
and $\xi^{\hati}$ is the standard timelike Killing vector of $AdS_{2p+1}$
(i.e. $\xi^{\hatt}=1, \xi^{\hata}=0$).

The ground state energy $E_0$  is also related to the Casimir energy of
a conformal field theory on $R \times S^{d-1}$. To see this
notice that the $R \times S^{d-1}$ is conformally related to Minkowski space.
One can thus obtain the vacuum energy on $R \times S^{d-1}$
by starting from Minkowski space where the expectation value
of the energy momentum vanishes and apply the conformal transformation
that maps it to $R \times S^{d-1}$. This would lead to a zero
vacuum energy if the transformations were non-anomalous, but because
in even dimensions there is a conformal anomaly one gets a non-zero
result. We refer to \cite{Cappelli:1988vw} for a discussion of the
$d=2$ and $d=4$ case. The fact that energy of $AdS_5$ is equal
to the Casimir energy of $N=4$ SYM was first discussed in
\cite{BK}.

\section{Other examples and global issues}

So far we have derived necessary conditions for the 
Witten-Nester energy to be well defined. 
Our discussion however was local in nature and thus
our conditions are certainly not sufficient.
In order to complete the analysis one has to
address global issues as well and establish 
the existence of Witten spinors with the 
asymptotics we discuss here. In this section we illustrate 
some of the subtleties by means of two examples. 

We assume in this paper that 
the boundary admits at least one spin structure that 
extends in the bulk. In general, however, the boundary 
manifold can admit many spin structures and only a subset 
of those may extend to the bulk\footnote{ 
A spin structure exists iff the second Steifel-Whitney
class vanishes, $0=w_2 \in H^2(M,Z_2)$, and the number of distinct
spin structures is equal to the dimension of $H^1(M,Z_2)$.
In particular, if $M$ is simply connected there is a unique 
spin structure.}. An elementary example 
that exemplifies the situation is the circle $S^1$. 
It admits two spin structures: spinors can be periodic
or anti-periodic around $S^1$. If a boundary $S^1$ is contractible in the 
interior then only the anti-periodic spinors extend, but 
if $S^1$ is not contractible both spin structures extend.
An example where such issues arise is in three dimensions
with boundary of topology $R \times S^1$. $AdS_3$ and the
BTZ black hole have a boundary of such topology, but in $AdS_3$ the 
circle is contractible in the interior whereas in the BTZ black hole
not. This is the first example we discuss below.
A related discussion for more general supersymmetric
spacetimes in $2+1$ AdS supergravity can be found in 
\cite{Izquierdo:1994jz}.

Another related issue is the question of regularity of the 
Witten spinor. One may successfully satisfy the local conditions 
that ensure finiteness of the Witten-Nester energy
by an appropriate choice of $\hat{\e}_{(-\half)}$, but
there may not exist a globally valid regular Witten spinor 
satisfying these boundary conditions. We illustrate this
issue with our second example, the AdS soliton. 

\subsection{Extremal BTZ Black Hole}

We discuss in the subsection the extremal BTZ black hole \cite{BTZ}.
The metric is given by 
\be
ds^2 = -N^2(\r) dt^2 + N^{-2}(\r) d\r^2 + \r^2 \left(d \f 
- \frac{\r_0^2}{\r^2} dt\right)^2
\ee
where 
\be \label{N}
N(\r)=\frac{1}{\r}(\r^2 - \r_0^2).
\ee
The spacetime has an extremal horizon at $\r=\r_0$ and a conformal 
boundary at $\r \to\infty$. 
Introducing a new radial coordinate
\be \label{rho}
\r= \sqrt{e^{2r}+\r_0^2}
\ee
we bring the metric in the form used in this paper
\be \label{BTZ_FG}
ds^2 = dr^2 + e^{2r} (- dt^2+d\f^2) + \r_0^2\, (dt - d\f)^2.
\ee
The horizon is now pushed to $r=-\infty$ and the boundary is at $r=\infty$.
This metric is of the general form (\ref{AdScoef})-(\ref{AdS_d2})
with $g_{(0)}$ the standard metric on $R \times S^1$.

The holographic stress energy tensor associated with this
solution can be computed using (\ref{tij}), 
\be
T_{tt} = T_{\f \f} = - T_{t \f} = \frac{\r_0^2}{\k^2},
\ee
where we used $K_{(2)ij}=-g_{(2)ij}$ and read off $g_{(2)}$ from
(\ref{BTZ_FG}).
The boundary metric has the timelike Killing vector
$\z_{(t)}=\z_{(t)}^i \pa_i = \pa/\pa t$ 
and the spacelike Killing vector $\z_{(\f)}=\z_{(\f)}^i \pa_i=\pa/\pa \f$
and we can use them to obtain the mass and angular momentum
of the solution,
\bea
M &=& - \int_0^{2 \p} d \f T^t_i \z_{(t)}^i 
= \int_0^{2 \p} d \f T_{tt} = \frac{\r_0^2}{4 G}, \\
J &=& - \int_0^{2 \p} d \f T^t_i \zeta_{(\f)}^i 
=\int_0^{2 \p} d \f T_{t\f} = - \frac{\r_0^2}{4 G}, 
\eea
where $G$ is Newton's constant,
so the metric is the extremal solution with $M=-J$. The extremal solution
with $M=J$ is given by the same metric but with $G_{t \f} \to - G_{t \f}$.
Setting $\r_0^2=0$ yields the massless solution. 

We now want to compute the Witten-Nester energy for the this solution.
The extremal BTZ black hole admits one Killing spinor \cite{Coussaert:1993jp},
and one could consider using it as a Witten spinor, as in our
discussion of AAdS spacetimes. We therefore need the 
explicit form of the Killing spinor.
The vielbein and spin connection of the metric (\ref{BTZ_FG})
are given by
\bea
&&E^{\hatt} = N(r)\,dt, \qquad E^{\hatr} = dr, \qquad 
E^{\hat{\f}} = \r(r) \,d\f - \frac{\r_0^2}{\r(r)} \, dt\;, \\
&&\o^{\hatr}{}_{\hat{\phi}} = - N(r) d \f, \qquad
\o^{\hat{\phi}}{}_{\hat{t}}=-\frac{\r_0^2}{\r(r)^2} dr, \qquad
\o^{\hatt}{}_{\hat{r}} = -\frac{\r_0^2}{\r(r)} d \f + \r(r) dt \nonumber 
\eea
where in these formulas $N$ and $\r$ are understood to be functions
of $r$ (cf (\ref{N}) and (\ref{rho})).
A straightforward computation shows that 
the Killing spinor is given by
\be \label{BTZ_Kil}
\e = \sqrt{N(r)}\, \hat{\e}_{(\mhalf)} \;,
\ee
where \(\hat{\e}_{(\mhalf)}\) is a constant spinor 
satisfying the following conditions
\be \label{BTZ_proj}
P^+\hat{\e}_{(\mhalf)} = P^-_{\f t} \hat{\e}_{(\mhalf)} =0 
%= - \G_{\hat{\f}\hatt}\hat{\e}_{(\mhalf)}
\ee
where $P^+=\half(1+\G^{\hatr})$ and $P^-_{\f t}=\half(1-\G^{\hat{\f}\hatt})$.
In (\ref{BTZ_proj}) we impose two projections on a two
dimensional spinor, so one might think that that there are no 
non-trivial solutions. In three dimensions however
there are two inequivalent representations of the 
gamma matrices:
(i) $\G^{\hatt}=i \s^2, \G^{\hat{\f}}=\s^1,
\G^{\hat{r}}=\s^3$, where $\s^k$ are the Pauli matrices,
and (ii)  $\Gamma'^{\hati}= -\G^{\hati}$.
In representation (i) 
we find that $\G^{\hatr}=-\G^{\hat{\f} t}$ 
and therefore $P^+=P^-_{\f t}$, so (\ref{BTZ_proj}) admits a 
non-trivial solution. Notice that the Killing spinor $\e$ is 
periodic in $\phi$, actually it is constant in $\f$, where the corresponding
AdS Killing spinor (\ref{killing}) is anti-periodic.

We now choose as a Witten spinor the Killing spinor (\ref{BTZ_Kil}).
The projection in (\ref{BTZ_proj}) implies 
$\G^{\hatt} \hat{\e}_{(-\half)}=\G^{\hat{\f}} \hat{\e}_{(-\half)}$
and this in turn implies that the boundary Killing vector,
\be 
\xi^{\hati}=\bar{\hat{\e}}_{(-\half)} \G^{\hat{i}} \hat{\e}_{(-\half)}
\ee
is a null Killing vector (since $\xi^{\hatt} = \xi^{\hat{\f}}$).
Choosing \(|\hat{\e}_{(\mhalf)}|^2 =1\) we have 
\be
\xi=\z_{(t)} + \z_{(\f)}.
\ee

Let us now compute the corresponding Witten-Nester conserved charge.
First we compute the ground state ``energy'',
\be
\half \hat{A}_{(1)} = 
-\hat{\e}_{(\mhalf)}^{\dag}\pa_{\f}^2 \hat{\e}_{(\mhalf)} = 0, 
\qquad \Rightarrow \qquad E_0=0,
\ee
since the spinor $\hat{\e}_{(\mhalf)}$ is constant. One should 
contrast this with the case of $AdS_3$, where $E_0=-\p/\k^2$.
We thus obtain,
\be
E_{WN}=- \int^{2\p}_{0}\,d\f\, T^t_i \xi^i = M+J=0,
\ee
as expected since the Witten-Nester energy
is by construction equal to zero for Witten spinors that are equal to 
Killing spinors. In other words, the Witten-Nester conserved charge
is a linear combination of the mass and angular momentum.

The Witten spinor, however, need not be equal to 
a Killing spinor. To obtain a Witten-Nester expression
for the mass we now consider the following spinor,
\be \e' = \sqrt{N(r)}\, \hat{\e}'_{(\mhalf)} \;,\ee
where
\be
P^+\hat{\e}'_{(\mhalf)} = P^+_{\f t} \hat{\e}'_{(\mhalf)} =0 
\ee
In order for this expression to admit a non-trivial solution
we must work with the  irreducible representation (ii) where $P^+=P^+_{\f t}$.

To show that this is a Witten spinor we compute,
\be
\nabla_{\hat{r}} \e' = \frac{\r_0^2}{\r^2} \e', \qquad 
\nabla_{\hat{\f}} \e' = \frac{\r_0^2}{\r^2} \G_{\hat{\f}}\e' 
\ee
from which we obtain
\be
(\G^{\hatr} \nabla_{\hat{r}} + \G^{\hat{\f}} \nabla_{\hat{\f}}) \e'=0.
\ee
This Witten spinor is associated with the null Killing vector field 
$\xi'^{\hat{i}} = \bar{\hat{\e}'}_{(\mhalf)} \G^{\hat{i}} 
\hat{\e}'_{(\mhalf)}$, where we normalize 
$|\hat{\e}'_{(\mhalf)}|^{2}¥=\half$,
\be
\x'^i\,\pa_i = \bar{\hat{\e}'}_{(\mhalf)} \G^{i} \hat{\e}'_{(\mhalf)}\,\pa_i 
= \half (\pa_{\hatt} -\pa_{\hat{\f}}) = \half (\z_{(t)} - \z_{(\f)}).
\ee
Let us now compute the Witten-Nester energy. The ground state energy $E_0$
is zero because $\hat{\e}'_{(\mhalf)}$ is constant and
\be
E'_{WN}= - \int^{2\p}_{0}\,d\f\, T^t_i \xi'^i = \half(M-J)=M.
\ee
Notice that the Witten spinor is regular for $\r^2 \geq \r_0^2$
or equivalently $r > -\infty$. Furthermore, a possible contribution to the 
Witten-Nester energy from the horizon vanishes since the 
Witten spinor vanishes at the horizon. 

This example illustrates a number of points. Firstly, we see explicitly the
dependence of the Witten-Nester construction on the spin structure 
and on the choice of Witten spinor. For $AdS_3$ one must choose
anti-periodic boundary conditions for the Witten spinor and the
dependence of $\hat{\e}_{(-\half)}$ on the $S^1$ coordinate $\f$ 
gives rise to the ground 
state energy $E_0$. In the BTZ case however the circle is not contractible and
periodic spinors are allowed. In fact one must choose periodic spinors 
if one wants to preserve supersymmetry. With this choice the ground state
energy vanishes. Another point that is illustrated by this example
is that one may have to consider Witten spinors that do not approach 
a Killing spinor asymptotically in order to obtain all conserved
charges.

\subsection{AdS Soliton}

In this subsection we discuss the AdS soliton \cite{Horowitz:1998ha}.
This solution has a toroidal boundary and negative energy 
but it has been conjectured 
\cite{Horowitz:1998ha} that it is the lowest energy solution 
within its asymptotic class. This was checked for small perturbations 
in \cite{Horowitz:1998ha}-\cite{Constable:1999gb} and additional 
support for this conjecture was presented in \cite{Galloway:2001uv}, \cite{anderson}.
The negative energy was shown in \cite{Horowitz:1998ha} to be 
(proportional to) the Casimir energy of $N=4$ SYM on $R \times T^3$.
In our general discussion we found that the Witten-Nester energy is 
equal to the holographic energy up to a ground state energy,
which is present in odd dimensions. 
This ground state energy for AAdS had the interpretation 
of Casimir energy for the dual CFT on $R \times S^{d-1}$. So one 
could have hoped that similar discussions would prove the positive
energy conjecture of \cite{Horowitz:1998ha}. However, inspection of the 
results in the literature and our results in section \ref{Hol_WN}
shows that this cannot be the case. The AdS soliton has negative
mass in all dimensions and this is incompatible with the bounds in 
section \ref{Hol_WN}. In particular, the mass of even dimensional AlAdS 
spacetimes is bounded by zero and of $AAdS_5$ by a positive quantity.
It will be instructive however to understand why our considerations
do not apply in this case.

The metric for the five-dimensional AdS soliton is given by 
\be
ds^2 =\frac{1}{N(\r)^2}\,\frac{d\r^2}{\r^2} 
+ \r^2\Bigl( \,N(\r)^2\,d\t^2-dt^2+dx^2+dy^2\,\Bigr),
%\;\;\;\;\;\t\in [0,\frac{\p}{a})
\ee
where 
\be N(\r) = \sqrt{1-\frac{a^4}{\r^4}}.
\ee
Regularity requires that $\t$ is identified with period $\frac{\p}{a}$, 
and we take $x, y$ to be periodic with periods $R_x, R_y$, respectively. 

A change of the radial coordinate, 
\be 
e^{r} = \frac{a^2}{\sqrt{2} \r \sqrt{1-N(\r)}} \;, \qquad {\rm or} \qquad
\r^2 = e^{2 r} + \frac{a^2}{4} e^{-2 r},
\ee
brings the metric in the form used in this paper,
\bea
ds^2&=& dr^2 + \left(e^{2r}+\frac{a^2}{4} e^{-2 r}\right) (-dt^2+dx^2+dy^2) 
+ \frac{(e^{2r}-\frac{a^2}{4} e^{-2 r})^2}{(e^{2r}+\frac{a^2}{4} e^{-2 r})}
d\t^2
\\ 
&=&dr^2 + e^{2r} \eta_{ij} dx^i dx^j
%(-dt^2+dx^2+dy^2+d\t^2) 
+ e^{-2r} \,\frac{a^4}{4} (-dt^2+dx^2+dy^2-3\,d\t^2) + \co(e^{-4r}) d\t^2.
\nonumber
\eea
{}From this metric we can read off the coefficients $g_{(n)}$ and 
obtain the coefficient $\hat{K}_{(n)}$ by using (\ref{k_g}).
Up to $n=4$ the only non-zero coefficients are
\be
\hat{K}_{(4)}{}^{\hat{t}}_{\;\;\hatt} 
= \hat{K}_{(4)}{}^{\hat{x}}_{\;\;\hat{x}}
=\hat{K}_{(4)}{}^{\hat{y}}_{\;\;\hat{y}} = \mhalf a^4 \qquad
\hat{K}_{(4)}{}^{\hat{\t}}_{\;\;\hat{\t}} = \frac{3}{2} a^4 
\ee
This boundary metric has a timelike Killing vector $\xi=\pa/\pa t$, 
and we can use (\ref{charges}) to compute the mass of the soliton,
\be \label{mass_sol}
E_h = \int d^3 x T_{tt} = - \frac{R_x R_y a^3}{16 G}
\ee
in agreement with \cite{Horowitz:1998ha}.

We now turn to the discussion of the Witten-Nester energy for 
this solution. Let us assume for the moment that the no-divergence 
condition in (\ref{pl_cond}) holds. The ground state energy $E_0$
is zero in this case since 
\be
\half\hat{A}_{(3)} 
= -\frac{1}{12} |\hat{K}_{(2)}{}^{\hatj}_{\;\;\hata} \G^{\hata}\G_{\hatj}
 \hat{\e}_{(\mhalf)} |^2 =0 
\ee
because $K_{(2)ij}=0$. To obtain the Witten-Nester energy,
we compute
\be
\hat{q}_{(3)} = \mhalf \hat{\e}_{(\mhalf)}^{\dag} 
\hat{K}^{\hatj}_{\;\;\hata(4)}  \G^{\hata}\G_{\hatj}\hat{\e}_{(\mhalf)} 
= -\frac{a^4}{4} |\hat{\e}_{(\mhalf)} |^2 <0,
\ee
so provided we can normalize $|\hat{\e}_{(\mhalf)} |^2=1$ we 
obtain 
\be
E_{WN}=E_h <0
\ee
which contradicts the positivity property of the Witten-Nester energy. 
 
Let us now discuss the no-divergence condition 
(\ref{pl_cond}) which for the solution at hand reads,
\be \label{soliton_condition1}
\hat{\e}_{(\mhalf)}^{\dag} (\,\pa_x^2+\pa_y^2+\pa_\t^2\,) \,\hat{\e}_{(\mhalf)}
+ c.c. = 0 \;.
\ee  
This condition is solved by a constant $\hat{\e}_{(-\half)}$ (which may be 
normalized to one) so our asymptotic conditions are satisfied.
Integrating the Witten equation with this boundary condition 
leads to 
\be \label{Wit_sol}
\e= \frac{a^2}{2}\,\frac{1}{\r^{3/2}} \, 
\frac{1}{\sqrt{N(\r)}}\,\sqrt{\frac{1+N(\r)}{1-N(\r)}}\hat{\e}_{(-\half)},
\ee
which is singular at $\r=a$. It follows that the step from 
(\ref{cons1}) to (\ref{cons2}) relating the manifestly positive 
bulk integral to a surface integral does not go through in this case.

One might have anticipated problems with regularity of the Witten
spinor since the circle corresponding to $\t$ is contractible in the 
interior, so the Witten spinor and in particular $\hat{\e}_{(-\half)}$
should be anti-periodic in $\t$. However, our 
$\hat{\e}_{(-\half)}$, and thus the Witten spinor in (\ref{Wit_sol}),
is periodic. If we demand that $\hat{\e}_{(-\half)}$ is antiperiodic
in $\t$, the condition (\ref{soliton_condition1})
cannot be satisfied (with $\hat{\e}_{(-\half)}$ periodic or anti-periodic
in $x,y$), and the Witten-Nester energy is not well-defined.

\section{Conclusions}

We derived in this paper conditions on the asymptotic structure 
of asymptotically locally AdS spacetimes such that 
their mass is bounded from 
below. This was done by computing a regulated version of the 
manifestly positive spinorial Witten-Nester energy and 
analyzing the condition for this energy to be finite.

The spinorial energy $E_{WN}$ is constructed from Witten spinors, 
i.e. spinor fields satisfying a Dirac-like equation on the 
initial-value hypersurface. It can be written either as
a bulk integral 
or as a surface integral at infinity.
The former is manifestly positive and the latter 
provides the connection with the conserved charges. 
The two expressions are equivalent, provided the 
Witten spinors are regular. For $AlAdS$ spacetimes,
the surface integral is not automatically finite, thus 
we introduce a cut-off $r$ in the radial 
direction to regulate the theory, as in previous work on 
holographic renormalization. 
The regulated $E_{WN}[r]$ can now be computed 
for general AlAdS spacetimes, provided that we know 
asymptotic solutions of the Witten equation.
We computed the most general asymptotic solutions 
of the Witten equation using methods similar to the 
ones in \cite{PS1}. As a technical 
remark, we note that the use of the formalism of \cite{PS1}
(instead of the near boundary expansion of \cite{FG,HS,dHSS}) 
was instrumental in allowing us to carry out this computation.
The coefficients of the asymptotic Witten spinors 
are determined locally (up to a specific order)
from the (still arbitrary at this stage) 
boundary value of the Witten spinor $\e_{(-\half)}$ and the 
boundary vielbein.

Having solved the Witten equation asymptotically, we then computed the 
regulated Witten-Nester energy. The expression involves a number
of local power-law divergences and 
a logarithmic divergence in odd dimensions. This means that 
not all AlAdS spacetimes possess a finite positive
Witten-Nester energy. The ones that do, have asymptotic 
data such that all divergences vanish identically.
Thus the vanishing of the divergences provides necessary 
conditions on the asymptotic data for 
the spacetime to possess a finite Witten-Nester energy.
The vanishing of the logarithmic divergence in odd dimensions
implies that the even dimensional conformal boundary should be a
conformally Einstein manifold. The number of power law divergences
depend on the spacetime dimension. In dimension three there are 
no power law divergences and in dimensions 4 and 5 there is one
such divergence. In these dimensions, the vanishing of the divergence
implies that the boundary manifold should admit a spinor 
satisfying a particular differential equation. It would be 
interesting to classify the four dimensional conformally 
Einstein spaces that admit such spinors. Such a list would provide
curved backgrounds for which ${\cal{N}}=4$ SYM is expected to be well defined.
Higher dimensions were only analyzed for 
AAdS spacetimes, i.e. for spacetimes that asymptotically approach
the exact AdS solution. In this case, all no-divergence conditions 
are satisfied if we take the Witten spinor to approach asymptotically
an AdS Killing spinor.

Having established the condition for finiteness we compared the 
finite part of the Witten-Nester energy with the expression of the 
holographic energy, $E_h$. In even dimensions the two agree exactly
and in odd dimensions they differ by a bounded quantity which only 
depends on the asymptotic data. We give an explicit expression 
of the bound for $AlAdS_3$, $AlAdS_5$ and discuss it for all AAdS spacetimes. 
A general feature is that it is negative in $4k-1$ dimensions 
and positive in $4k+1$ dimensions ($k{=}1,2,\ldots$). This difference 
between $E_{WN}$ and $E_h$ in odd dimensions is 
due to the fact that $E_{WN}$ is by construction equal to zero 
for supersymmetric solutions, while the holographic energy may not be 
zero because of the conformal anomaly. 

In this paper we only analyzed local properties that 
follow from the asymptotic analysis. In order to rigorously
establish the bounds, one has to show 
existence of Witten spinors with the asymptotics we discuss. 
It is clear from examples that such a discussion will depend 
sensitively on global properties. For example,
one would have to understand the dependence of the 
construction on spin structures.
To illustrate such subtleties
we discussed two examples, the extremal BTZ black hole and the AdS soliton.
The extremal BTZ and $AdS_3$ spacetimes 
have the same conformal boundary, but the Killing
spinors are periodic (along the compact boundary direction) 
in the supersymmetric BTZ case
and antiperiodic in the case of $AdS_3$. The energy bound in these cases thus
depends on the spin structure. The example of the BTZ black hole 
also illustrates the fact that,  in order to construct 
all conserved charges, it may 
be necessary in some cases to consider Witten spinors
that do not approach asymptotically bulk Killing spinors. 
The AdS soliton gives an example where all 
local requirements can be satisfied, but a global regular 
Witten spinor with these boundary conditions does not exist.

In this paper we have restricted our attention to the case of pure gravity,
but the discussion can be generalized to include matter. This is interesting
both intrinsically and also from the point of view of the AdS/CFT 
correspondence.  A particularly interesting case is that 
of domain wall backgrounds since they are dual to holographic RG flows. 
A stability analysis for a class of such spacetimes was presented in 
\cite{Townsend:1984iu,Skenderis:1999mm,Freedman:2003ax}.
An extension of our analysis in this direction will  
lead to a systematic search for stable backgrounds
supported by matter fields.

\section*{Acknowledgments}
We would like to thank D. Freedman, C. N\'{u}\~{n}ez and M. Schnabl
for discussions. MC would like to thank A. Strominger and the 
Center for Mathematical Sciences in Zhejiang university, 
where part of this work was completed, for the hospitality. 
KS is supported by NWO and MC by FOM.

\appendix

\section{Conventions and Notations} \label{conventions}
\setcounter{equation}{0}  % reset counter
Our index conventions are as follows
\be
{\{\m\}= \{r,\{i\}\}\;\;;\;\{i\}=  \{t,\{a\}\}\;\;;\;\{a\}=\{1,2,....,d-1\}
\;\;;\;\;\{\a\} = \{r,\{a\}\}}\;,\\
\ee
and hatted indices stand for flat indices. We use mostly plus signature.
Our spinor conventions and covariant derivatives are given by  
\bea
{\bare=\dage\,\G_{\hat{t}}}\\
{\{\G^{\m},\G^{\n}\}=2\,G^{\m\n}}\\
{(\Ga)^{\dag}=\Ga\;\;\; ,(\Gr)^{\dag}=\Gr\;\;\; ,(\Gt)^{\dag}=-\Gt}\\
{\cal D}_{\m}\e=(\pa_{\m}+\frac{1}{4}\O_{\m}^{\,\hatn\hat{\r}}\,
\G_{\hatn\hat{\r}})\e \\
{\cal D}_{[\m}\,{\cal D}_{\n]}\e = \frac{1}{8}\,R_{\m\n\r\d}\,
 \G^{\r\d}\,\e\\
 {\na_{\m}\e \equiv ({\cal D}_{\m}+\frac{1}{2}\G_{\m}¥)\e}.
\eea

\section{The radial and the time slices} \label{slices}
\setcounter{equation}{0}  % reset counter

We list here the various slices used in the main text.
We consider an AlAdS spacetime $M$ with conformal boundary $\pa M$.
As discussed in the main text, we can always choose coordinates
near the boundary where the metric looks like
\bea \label{bulkmetric}
ds^2&=& G_{\m\n}\,dx^{\m}dx^{\n} = dr^2 + \g_{ij}(x,r) dx^i dx^j \nonumber \\
&=& E^{\hatr} \otimes E^{\hatr}
+ \sum_{\hati,\hatj} \eta_{\hati \hatj}\,E^{\hati} \otimes E^{\hatj}\;,
\eea
where $\eta_{ij}$ is the Minkowski metric in $d$ dimensions and
we introduce the vielbein 1-forms
\be
E^{\hat{\mu}} = E^{\hat{\m}}_\mu dx^\mu.
\ee
The choice of coordinates in (\ref{bulkmetric}) implies that we
can choose
\be
E_r^{\hat{r}}= 1 + \co(e^{-(d-1)r/2}), \quad
E^r_{\hat{r}} =1 + \co(e^{-(d-1)r/2}), \quad
E_{r}^{\hat{i}}=\co(e^{-(d-1)r/2}), \quad
E_{\hat{r}}^{i}=\co(e^{-(d-1)r/2}),
\ee
which implies
\be
E^{\hat{r}}=d r, \qquad \vec{\pa}_{\hat{r}} \equiv E^\mu_{\hat{r}} \pa_\mu
= \pa_r.
\ee
up to the order indicated above.

\begin{figure}
\begin{center}
\scalebox{1}{\rotatebox{-0}{\includegraphics{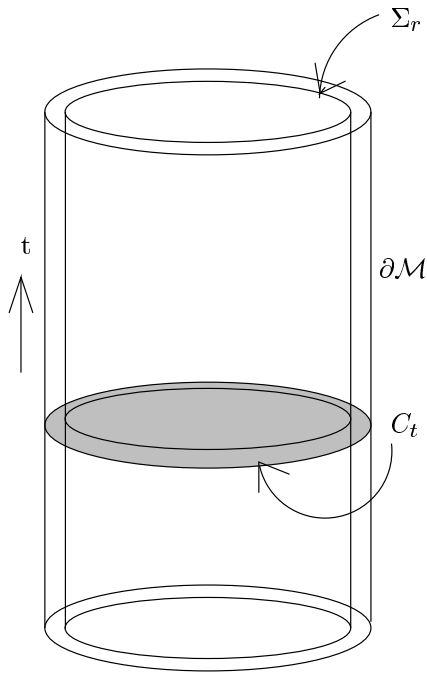}}}
\end{center}
\caption[]{\small We illustrate in this figure the various
slices used in this paper. The shaded area is the initial value
surface, $\pa M$ is the conformal boundary of the spacetime 
and $\S_r$ is the regulated surface.}
\label{cauchy}
\end{figure} 

\subsection*{The radial slice \(\S_r\)}

The radial slice is defined by its normal $E^{\hat{r}}$. With the
coordinate choice in (\ref{bulkmetric}), this is the $r=const$ slice.
The induced metric is given by
\be \label{rslice}
ds^2_r = \g_{ij}(r,x) \,dx^{i}dx^j =
\sum_{\hati,\hatj} \eta_{\hati \hatj}\,E^{\hati} \otimes E^{\hatj}
\ee
As $r \to \infty$, $\S_r$ approaches the conformal boundary $\pa
M$. In this limit the induced metric $\g_{ij}(x,r)$ blows up and
only a conformal structure is well-defined. One can pick a
specific representative $g_{(0)}$ by a specific choice of defining function
$\r$ (a defining function is a positive function that has a single
zero at the boundary). Choosing as defining function $\r=e^{-r}$
we get as a boundary metric 
\be \label{bdrymetric3} 
ds_{\infty}^2 = g_{(0)ij} \,dx^i dx^j = \sum_{\hati, \hatj} 
\eta_{\hati \hatj}\; e^{\hati}
\otimes e^{\hatj} \ee where \(e^{\hati}= \lim_{r \to \infty}
e^{-r} E^{\hati}\).

\subsection*{The time slice \(C_t\)}

We consider the time slice $C_t$ defined by its normal $E^{\hat{t}}$.
The induced metric on $C_t$ is
\bea
ds^2_{t} &\equiv& {}^tG_{\m\n} d x^{\m}d x^{\n}
=\left(G_{\m\n} + E^{\hatt}_{\m}E^{\hatt}_{\n}\right)d x^{\m}d x^{\n}
\nonumber \\
&=&\sum_{\a} E^{\hat{\a}} \otimes E^{\hat{\a}} 
=E^{\hat{r}} \otimes E^{\hat{r}}+
\sum_{a} E^{\hat{a}} \otimes E^{\hat{a}}.
\eea

\subsection*{The boundary of the time slice \(C_t \cap \S_r \)}

The induced metric on the intersection of the two slices
is
\be
ds^2_{(rt)} \equiv {}^t\g_{ij} dx^i dx^j 
=\Bigl(\,\g_{ij}+E^{\hatt}_i E^{\hatt}_j\,\Bigr)\,dx^i dx^j
= \sum_{a} E^{\hata} \otimes E^{\hata} \\ \nonumber
\ee
As $r \to \infty$, and with the same defining function as before,
we get for the metric on \(C_t \cap \pa M \)
\be
ds^2_{\infty,t} \equiv {}^tg_{(0)ij} dx^i dx^j
=(\,g_{(0)ij}+e^{\hatt}_i e^{\hatt}_j\,)dx^i dx^j,
\ee
where \(e^{\hatj}_i= \lim_{r \to \inf}
(e^{-r}\,E^{\hatj}_i)\) is the vielbein of the boundary metric \(g_{(0)ij}\).

\section{Asymptotic expansions} \label{as_exp}

In this appendix we present some of the technical details
needed in order to obtain the asymptotic solution of the Witten
spinor.

The dilatation operator (\ref{dil}) is given
in terms of the vielbein by
\be
\d_D = \int d^{d}x\,E^{\hatj}_{i}\,\frac{\d}{\d E^{\hatj}_{i} }\, ,
\ee
and the second fundamental form and radial derivative admits the expansions
\bea
K^{i}_{\,j}[\g] &=& \d^{i}_{\,j}
+ \sum_{k=1}^{[\frac{d-1}{2}]} K_{(2k)}{}^{i}_{j}+
K_{(d)}{}^{i}_{j}
+\til{K}_{(d)}{}^{i}_{j} (-2r) + \cdots \\
\pa_r &=& \d_D +
\sum_{k=1}^{[\frac{d-1}{2}]}\pa_{r(2k)} + \pa_{r(d)}-(2 r) \tilde{\pa}_{r(d)}
+ \cdots \nonumber \\
\pa_{r(n)} &=& \int d^d x K_{(n) ij} \frac{\d}{\d \g_{ij}},
\qquad \tilde{\pa}_{r(d)} = \int d^d x \til{K}_{(d) ij} \frac{\d}{\d \g_{ij}},
\eea
where $[\frac{d-1}{2}]$ 
denotes the integer part of $\frac{d-1}{2}$ and $\til{K}_{(d)ij}$ is
zero when $d$ is odd. Furthermore, since
$K_{ij}=\dot{E}_{(i}^{\hati} E_{j)}^{\hatj} \eta_{\hati \hatj}$,
$\dot{E}_{i}^{\hati}$ admits an expansion of the form
\be
\pa_r  E_i^{\hat{j}} = E_i^{\hatj} + \sum_{k=0}^{d-1}
\dot{E}_{(k)}{}^{\hatj}_{i}+
(-2r) \dot{\til{E}}_{(d-1)}{}^{\hatj}_{i} +\cdots
\ee

In our coordinate system, the spin connections are given by
\be
\Omega_r^{\hati \hatj} = E^{k[\hati}\pa_{r}E^{\hatj]}_{k}, \qquad
\Omega_i^{\hatj \hatr} = E^{k\hatj} K_{ik}, \qquad \Omega_i^{\hati \hatj},
\ee
and the covariant derivatives take the form,
\bea
\na_{r}\e&=&(\,\pa_{r}+\frac{1}{4}\,
E^{k[\hati}\pa_{r}E^{\hatj]}_{k}\,
\G_{\hati\hatj}+\frac{1}{2} \G_r \,)\,\e\,, \nonumber \\
\na_{i}\e 
&=& (\,D_i + \half  E^{k\hatj} K_{ik} \G_{\hatj\hatrr}
+\frac{1}{2} \G_i\,)\,\e\;. \label{cov}
\eea
where $D_i$ is the covariant derivative of the induced metric $\g_{ij}$.

Using the results above one can work out the asymptotic
expansion of the covariant derivatives and the operator
$\sd \equiv \G^{\hatr} \na_{\hatr} + \G^{\hata} \na_{\hata}$
that appear in the Witten equation,
\bea
\sd_{(0)}&=&(\dil\,+\frac{d-1}{2})\Gr +\frac{d}{2} \nonumber \\
\sd_{(1)}&=&\G^{\hata}D_{\hata} \\
\sd_{(2n)}&=& \Bigl( 2\,\int d^{d}x\,\g_{ik}\,
K_{(2n)}{}^{i}_{\;\;j}\,\frac{\d}{\d \g_{jk}}
+ \frac{1}{4} \,E^{k[\hati} \dot{E}_{(2n-1)}{}^{\hatj]}_{k} \,\G_{\hati\hatj}
+\frac{1}{2}\,K_{(2n)}{}^{\hatj}_{\;\hata}\,\G^{\hata}\G_{\hatj}
\Bigr)\,\Gr \nonumber \\
\sd_{(d)} &=&
\Bigl( 2\,\int d^{d}x\,\g_{ik}\,K_{(d)}{}^{i}_{\;\;j}\,\frac{\d}{\d \g_{jk}}
+ \frac{1}{4} \,E^{k[\hati} \dot{E}_{(d-1)}{}^{\hatj]}_{k} \,\G_{\hati\hatj}
+\frac{1}{2}\,K_{(d)}{}^{\hatj}_{\;\hata}\,\G^{\hata}\G_{\hatj}
\Bigr)\,\Gr \nonumber \\
\tilde{\sd}_{(d)}&=&
\Bigl( 2\,\int d^{d}x\,\g_{ik}\,\til{K}_{(d)}{}^{i}_{\;\;j}\,
\frac{\d}{\d \g_{jk}}
+ \frac{1}{4} \,
E^{k[\hati} \dot{\til{E}}_{(d-1)}{}^{\hatj]}_{k} \,\G_{\hati\hatj}
+\frac{1}{2}\,\til{K}_{(d)}{}^{\hatj}_{\;\hata}\,\G^{\hata}\G_{\hatj}
\Bigr)\,\Gr  \;. \nonumber
\eea
where $ n=1,2,....,[\frac{d-1}{2}]$.
Note also that $\d_D E^{k\hati} =- E^{k\hati}$, so 
$E^{k\hati}\equiv E^{k\hati}_{(1)}$ and that
$\tilde{\sd}_{(2k+1)}=0$.

Observe that
\bea
&& \left[ \til{\sd}_{(d)}, P^\pm \right] = 0,
\qquad \left[\sd_{(k)}, P^\pm \right] =0 \ (k\neq1),
\qquad \sd_{(1)} P^{\pm} = P^{\mp} \sd_{(1)} \nonumber  \\
&&\sd_{(0)}\Pm = (-\dil+\frac{1}{2}) \Pm, \qquad
\sd_{(0)}\Pp = (\dil + d -\frac{1}{2})\Pp \label{begin}
\eea

We will also use in the main text the asymptotic expansion of the
operator \(\G^{\hata}\na_{\hata}\). It is given by
\bea
\G^{\hata}\na_{\hata(0)}&=&\frac{d-1}{2}\Gr+\frac{d-1}{2}=(d-1)\Pp
\nonumber \\
\G^{\hata}\na_{\hata(1)}&=&\G^{\hata}D_{\hata} \label{d_a_begin}  \\
\G^{\hata}\na_{\hata(2n)}&=&\frac{1}{2}\,
K^{\hatj}_{\;\hata(2n)}\G^{\hata}\G_{\hatj}\Gr
\;\;,\;\; n=1,2,....,[\frac{d-1}{2}] \nonumber \\
\G^{\hata}\na_{\hata(d)}&=&\frac{1}{2}\,K^{\hatj}_{\;\hata(d)}
\G^{\hata}\G_{\hatj}\Gr \nonumber \\
\G^{\hata}\tilde{\na}_{\hata(d)}
&=&\frac{1}{2}\,\tilde{K}^{\hatj}_{\;\hata(d)}
\G^{\hata}\G_{\hatj}\Gr\; .\nonumber
\eea

\section{Properties under Weyl transformations} \label{app_weyl}

We discuss in this appendix the Weyl transformation properties
of the conditions of the absence of divergences. We focus on the
Weyl transformation properties of (the integral of) $q_{(1)}$
but the discussion can be extend to the other coefficients.

We wish to know how the leading divergence
\bea \label{q1}
&&\int_{C_t \cap \S_r} \sqrt{{}^t\g}\ q_{(1)}= 
\int d^{d{-}1}\! x \sqrt{{}^tg_{(0)}}\ e^{(d-2) r} \hat{q}_{(1)}
(1 + \co(e^{-2 r}))  \\
&&=-\int d^{d{-}1}\! x \sqrt{{}^tg_{(0)}}\ e^{(d-2) r}
\hat{\e}_{(-\half)}^\dagger
\left(\frac{1}{(d-1)} (\G^{\hata} \hat{D}_{\hata})^2
+ \half \hat{K}_{(2)}{}^{\hatj}_{\hata}
\G^{\hat{a}} \G_{\hatj}\right) \nonumber
\hat{\e}_{(-\half)} \left(1 + \co(e^{-2 r})\right)
\eea
transforms under a local Weyl transformation,
\be \label{new_g0}
\bar{g}_{(0)ij} = e^{2\s(x)}g_{(0){ij}}\;.
\ee
(The hat notation is defined in (\ref{hat})).
In order $\bar{g}_{(0)ij}$ to admit a timelike
Killing vector we need to impose
the following condition on \(\s(x)\):
\be \label{sigma_condition1}
\s_{,\hatt} \equiv e^{i}_{\hatt}\pa_{i}\s=0\;,
\ee

The Weyl transformation of (\ref{q1}) can be worked out using
\bea
\bar{{\hat{D}}}_{\hata} &=& e^{-\s}\Bigl(\,{\hat{D}}_{\hata}+\half
\s_{, \hatb}\,\G_{\hata}^{\;\;\hatb}\Bigr)\\
\bar{\hat{K}}_{(2)ij} &=& \hat{K}_{(2)ij} - \hat{D}_i \hat{D}_j \s
+ \hat{D}_i \s\hat{D}_j \s - \half (\hat{D} \s)^2 g_{(0)ij} \nonumber \\
\bar{\hat{\e}}_{(-\half)} &=& e^{\s/2} \hat{\e}_{(-\half)}.
\nonumber \eea 
Using these results we derive,
\be \label{weyl} \int d^{d{-}1}\! x \sqrt{{}^tg_{(0)}}\ e^{(d-2)
r} \hat{q}_{(1)} =\int  d^{d{-}1}\! x \sqrt{{}^t\bar{g}_{(0)}}\
e^{(d-2) (r - \s)} \left( \bar{\hat{q}}_{(1)} -
\bar{\hat{\e}}_{(-\half)}^\dagger \s_{,\hatb} \G^{\hatb \hata}
\bar{\hat{D}}_{\hata} \bar{\hat{\e}}_{(-\half)}\right) \ee 
The overall factor is due to the dilatation transformation property of
$q_{(1)}$. 
We will now show that the additive term is required by
the invariance of the Witten-Nester energy under diffeomorphisms.

Weyl transformations on the boundary are induced by special bulk
diffeomorphisms, 
\be e^{\bar{r} + \s(\bar{x})} = e^{r}\,(\,1+{\cal{O}}(e^{-2r})\,), 
\qquad\ \bar{x}^{i} = {x^{i}}
\,(\,1+{\cal{O}}(e^{-2r})\,)\;. 
\ee 
The regulated Witten-Nester
energy is invariant under this transformation provided we also
transform the cut-off, \be E_{NW}[r] = \bar{E}_{NW}[\bar{r}
+\s(\bar{x})]. \ee The normal to the surface $\bar{r} +\s(\bar{x})$
is given by $l_\mu = {\cal D}_\mu (\bar{r} + \s(\bar{x}))$.
Inserting this in the definition of the Witten-Nester energy and
considering the leading term in the limit $\bar{r} \to \infty$ we
get \be \bar{E}_{NW}[\bar{r} +\s(\bar{x})] \sim \int d^{d{-}1}\! x
\sqrt{{}^t\bar{g}_{(0)}}\ e^{(d-2) \bar{r}} \left(
\bar{\hat{q}}_{(1)} - \bar{\hat{\e}}_{(-\half)}^\dagger \s_{,\hatb}
\G^{\hatb \hata} \bar{\hat{D}}_{\hata} \bar{\hat{\e}}_{(-\half)}+
c.c.\right) \ee which agrees with the rhs of (\ref{weyl}). The
additive term is due to $\s$ dependence of the normal vector
$l_\mu$.

\section{Bulk Killing spinors and Witten spinors} \label{Kil_Wit}

We show in this appendix that for an AlAdS spacetime, the
bulk Killing spinor $\e$ admits
the asymptotic expansion
\be
\e=\e_{(\mhalf)}+\e_{(\half)}+ \cdots
\ee
with
\be
\hat{\e}_{(\mhalf)} = \Pm  \hat{\e}_{(\mhalf)}, \quad
\hat{\e}_{(\half)} = - \G^i {\hat{D}}_{i} \hat{\e}_{(\mhalf)}
\quad \forall \ i, \ {\rm no\ sum\ over}\ i
\ee
Furthermore,
\be\x^{i}= \bar{\hat{\e}}_{(\mhalf)}\G^{i}\hat{\e}_{(\mhalf)}\ee
is a timelike or null boundary conformal Killing vector. \newline

{\it{Proof:}}\newline
The asymptotic expansion of the covariant derivatives can be obtained
from the results in appendix \ref{as_exp}. Starting from (\ref{cov})
we obtain
\bea
\na_{r}\e &=&
\Bigl( (\d_{D}+\half \G_{\hatr}) + \na_{(2)}+ \cdots\Bigr) \;\e \\
\na_{i} \e  \label{dr}
&=& (E^{\hatj}_{i} \G_{\hatj} \Pp + D_{i} +\na_{(2)i} + \cdots )\;\e\;.
\label{di}
\eea
We can now solve asymptotically the Killing spinor equations
\be \na_{\m}\e=0.
\ee
The radial equation, $\na_r \e=0$, implies that
\bea
\e&=& e^{\half r} \hat{\e}_{(\mhalf)}+e^{\mhalf r} \hat{\e}_{(\half)}
+ \cdots\\
 \hat{\e}_{(\mhalf)} &=& \Pm  \hat{\e}_{(\mhalf)} \;\;;\;\;
 \hat{\e}_{(\half)} = \Pp  \hat{\e}_{(\half)}
\eea
Inserting this in the spatial equations, $\na_i \e=0$,  we obtain
\be \label{killing1}
{\hat{D}}_{i} \hat{\e}_{(\mhalf)} + \G_{i} \hat{\e}_{(\half)} = 0\;,
\qquad \Leftrightarrow  \qquad
\hat{\e}_{(\half)} = - \G^i {\hat{D}}_{i} \hat{\e}_{(\mhalf)}
\quad \forall \ i \ ({\rm no\ sum})
\ee
where \({\hat{D}}_{i} \) and \(\G_{i} \equiv e^{\hatj}_{i} \G_{\hatj}\) are
defined with respect to the boundary metric \(g_{(0)ij}\).

Notice that an asymptotic Killing spinor is in particular a Witten
spinor, but not vice versa. For instance, the sub-leading asymptotic
coefficient of a Witten spinor is given by (\ref{sol1}),
\be
\hat{\e}_{(\half)}
= -\frac{1}{d-1} \G^{\hata}
{\hat{D}}_{\hata} \hat{\e}_{(\mhalf)}\;,  \qquad {\rm Witten\ spinor}
\ee
Unless $\G^{\hata} \hat{D}_{\hata} \hat{\e}_{(-\half)}
=\G^{\hatb} \hat{D}_{\hatb} \hat{\e}_{(-\half)}$, for all
$a$ and $b$, the Witten spinor will not asymptote to a Killing
spinor.

The fact that $\xi^i = \bar{\hat{\e}}_{(\mhalf)}\G^{i}\hat{\e}_{(\mhalf)}$
is a conformal Killing vector follows by direct computation using
the asymptotics of the Killing spinor,
\bea\nonumber
{\hat{D}}_{(i}\,\x_{j)} &=& {\hat{D}}_{(i}\,\left(\bar{\hat{\e}}_{(\mhalf)}
\G_{j)} \hat{\e}_{(\mhalf)}\right)\\ \nonumber
%&=&\,\overline{ {\hat{D}}_{(i}\hat{\e}_{(\mhalf)}} \;
%\G_{j)} \hat{\e}_{(\mhalf)} + \bar{\hat{\e}}_{(\mhalf)}
%\G_{(j} {\hat{D}}_{i)}\hat{\e}_{(\mhalf)}\\ \nonumber
&=&\bar{\hat{\e}}_{(\half)}\G_{(i}
\G_{j)} \hat{\e}_{(\mhalf)}-\bar{\hat{\e}}_{(\mhalf)}\G_{(j}
\G_{i)} \hat{\e}_{(\half)}\\ \nonumber
&=& g_{(0)ij} \, (\,\bar{\hat{\e}}_{(\half)}\hat{\e}_{(\mhalf)}
-\bar{\hat{\e}}_{(\mhalf)}\hat{\e}_{(\half)}\,)\\
&=& \frac{1}{d}  g_{(0)ij} {\hat{D}}_{k} \,
(\bar{\hat{\e}}_{(\mhalf)}
\G^{k} \hat{\e}_{(\mhalf)}) \nonumber \\
&=& \frac{1}{d}  g_{(0)ij} \,{\hat{D}}_{k} \x^{k}.
\eea

We now prove that $\xi^{\hati} = 
\bar{\hat{\e}}_{(\mhalf)}\G^{\hati}\hat{\e}_{(\mhalf)}$ 
is timelike or null. Let us introduce the hermitian matrix 
\be
A = v_{\hat{a}} \G_{\hat{t}} \G^{\hat{a}}, \qquad 
v^2 \equiv \eta^{\hat{a} \hat{b}} v_{\hat{a}}  v_{\hat{b}} =1, 
\qquad i=\{t, \{a\}\}.
\ee
and consider $\hat{\e}_{(\mhalf)}$ that is an eigenvector of $A$,
\be \label{a_eigen}
A \hat{\e}_{(\mhalf)} = a \hat{\e}_{(\mhalf)}.
\ee
Since $A^2=1, a^2=1$ too. Multiplying (\ref{a_eigen}) by 
$\hat{\e}_{(\mhalf)}^\dagger$ and squaring we
get
\be
(\xi^{\hat{0}})^2 = \left( \sum_{\hata} v_{\hat{a}} \xi^{\hat{a}} \right)^2.
\ee
Elementary algebra shows that
\be
\sum_{\hat{a} < \hatb} (\xi^{\hata} v^{\hatb} - \xi^{\hatb} v^{\hata})^2
=\sum_{\hata} (\xi^{\hata})^2 
- \left( \sum_{\hata} v_{\hat{a}} \xi^{\hat{a}} \right)^2 
\geq 0,
\ee
which implies
\be
|\xi|^2 \equiv - (\xi^{\hat0})^2 + \sum_{\hata} (\xi^{\hata})^2 \leq 0.
\ee

\section{Killing Spinors of AdS${}_{d+1}$ in global coordinates} \label{AdSK}
\setcounter{equation}{0}  % reset counter

We discuss in this appendix the structure of the Killing spinors
for AdS\(_{d+1}\) spacetimes in coordinates 
\be 
ds^2= dr^2  -N_+^{2}(r) \,dt^2 + N_-^{2}(r)\, d\Omega_{d-1}^2\;,
 \ee
where 
\be N_\pm(r) = e^r \pm \frac{1}{4} e^{-r}  
\ee and
\(d\Omega_{d-1}^2\) is the standard metric on \(S^{d-1}\),
\be\label{sphere_metric} 
d\Omega_n^2= d\th_n^2+\sin^2\th_n\,d\Omega_{n-1}^2\;\;;\;\;
d\Omega_1^2= d\th_1^2 \;,\ee 
The radial coordinate $\r$ usually  used in the  standard global coordinates
is given by $\r=N_-$.

We find that the Killing spinors can be written in the
following compact form
\be\label{killing} 
\epsilon= e^{\frac{r}{2}} \,\ei
+e^{-\frac{r}{2}} \,\eii \;,
\ee 
where 
\bea \ei &=&\Pm
{\cal{O}}^+_{d-1} {\cal{O}}_{d-2} ... {\cal{O}}_{1} {\cal{O}}_{t}
\eta \label{eta} \\ 
\eii &=&-\frac{1}{2} \Pp {\cal{O}}^-_{d-1} {\cal{O}}_{d-2}
... {\cal{O}}_{1} {\cal{O}}_{t} \eta \;, \nonumber
\eea 
with \(\eta\) a constant spinor and
\bea {\cal{O}}_{t}&=&
e^{-\frac{t}{2}\G^{\hat{t}}} = \cos\frac{t}{2}
-\sin\frac{t}{2}\,\G^{\hat{t}} \\
{\cal{O}}_{j}&=& e^{\frac{\th_j}{2}\G^{\widehat{j{+}1},\hat{j}}} =\cos\frac{\th_j}{2}
+ \sin\frac{\th_j}{2}\,\G^{\widehat{j{+}1},\hat{j}} \;\;\;\;\;j=1,..,d-2 
\nonumber \\
{\cal{O}}^\pm_{d-1} &=& \cos\frac{\th_{d-1}}{2}
\pm\sin\frac{\th_{d-1}}{2}\,\G^{\widehat{d{-}1}}.  \nonumber 
\eea 

\begin{flushleft}
{\it Proof}
\end{flushleft}

The covariant derivatives are given by,
\bea 
\na_{\hat{r}} &=& \pa_r +\half \G_{\hat{r}} \label{nabla_k} \\ 
\na_{\hat{t}}&=& \frac{1}{N_+}\,(\,\pa_t +
e^r \G_{\hat{t}}\Pp+\frac{1}{4} e^{-r} \G_{\hat{t}}\Pm \,) 
\nonumber \\
\na_{\hat{\th}_k} &=& \frac{1}{N_-}\,
(\,\hat{D}_{\hat{k}} + e^r \G_{\hat{k}}\Pp-\frac{1}{4} e^{-r}
\G_{\hat{k}}\Pm \,),\;\;\;\;\;k=1,\ldots,d-1 \nonumber 
\eea 
where \(\hat{D}_{\hat{k}} =e^{k}_{\hat{k}} \hat{D}_k\) denotes the 
covariant derivatives on the unit sphere. 
Explicitly,
\bea\hat{D}_k &=& \frac{\pa}{\pa \th_k} +
\frac{1}{4}\, \omega_i^{\;\;kl} \G_{kl} \label{D_sph} \\ 
\omega_i^{\;\;kl}&=& \delta^{k}_i\,\cos \th_l \prod^{l-1}_{m=i+1}
\,\sin \th_m \qquad (k < l)\;, \nonumber  \\
e^{\hat{k}}_k &=& \prod^{d-1}_{m=k+1} \,\sin \th_m. \nonumber
\eea 
 
Using these expressions, one can easily verify that
(\ref{killing}) satisfy \(\na_{\hat{r}}\epsilon= \na_{\hat{t}}
\epsilon =0 \), so we concentrate on the spherical part. From 
(\ref{nabla_k}) we see that $\na_{\hat{\th_k}} \e =0$ is
equivalent to the following equations,
\bea \label{new_nablai}
\hat{D}_{\hat{k}} \ei+ \G_{\hat{k}}\,\eii &=& 0\;,\\
\nonumber  
\hat{D}_{\hat{k}} \eii -\frac{1}{4}\,\G_{\hat{k}}\,\ei &=& 0\;.
\eea 
These equations are easily shown to hold for \(k=d-1\), so in the
following we discuss the cases \(k=1,2,.., d-2\).  Our
proof is similar in spirit with the discussion in \cite{Lu:1998nu}.

We begin with the fact\footnote{To avoid cumbersome notation
we drop the "hats" from the indices of the gamma matrices
in the rest of this appendix.} 
 \be 
\pa_j\calO_k= \delta_{jk}\,\half \G^{j+1,j} \calO_j, 
\ee 
which can be used to rewrite
(\ref{new_nablai}) as
\be\label{new_new_nablai} 
\calO^{\pm}_{d-1}\,U_j^{\;\;d-2} +
\o_j^{\;\;jl}\,\G_{jl} \calO^{\pm}_{d-1} - e^{\hatj}_j \G_j
\calO^{\mp}_{d-1} = 0, 
\ee
where \(U_j^{\;\;k}\;\;(k \leq j)\) is defined by
\be U_j^{\;\;k} = \calO_k\, U_j^{\;\;k-1}
\calO^{-1}_k; \;\;\;\;\;\;U_j^{\;\;j} \equiv \G_{j+1,j} \;.
\ee
Equations (\ref{new_new_nablai}) can further be rewritten as
\be \label{neweqn}
\calO_{d-1}^\pm \,U_j^{\;\;d-2} \calO_{d-1}^\mp +
\cos\th_{d-1}\,\o_j^{\;\;jl}\,\G_{jl} \mp e^{\hatj}_j\,\G_j
\,(1\mp \sin\th_{d-1}\, \G_{d-1}) = 0\;,
\ee
where we have used the relations
\bea
\calO_{d-1}^+\,\calO_{d-1}^- &=&
\calO_{d-1}^-\,\calO_{d-1}^+ = \cos \th_{d-1} \\
(\calO_{d-1}^\pm)^2 &=& 1 \pm \sin \th_{d-1}\,\G_{d-1} \;. \nonumber 
\eea
Using (\ref{D_sph}) and the relation
\be 
\calO^\pm_{d-1} \,\G_{j,d-1}\,
\calO^\mp_{d-1} = \G_{j,d-1}\mp \sin\th_{d-1} \G_j,
\ee
one can prove (\ref{neweqn}) provided $U_j{}^k$ is given by
\be \label{Uexp}
\label{lemma} U_j^{\;\;k} = - \sum^{k}_{l>j} \o_j^{\;\;jl}
\,\G_{jl} - \sec\th_{k+1}\, \o_j^{\;\;j,k+1} \,\G_{j,k+1}.
\ee
We now prove this relation by
induction. First observe that \(U_j^{\;\;j}=\G_{j+1,j}\)
satisfies (\ref{lemma}). Suppose now that (\ref{lemma}) is satisfied
for \(U_j^{\;\;k}\) for some \(k< d-2\). Using 
\be \calO_{k+1}
\,\G_{j,k+1} \calO_{k+1}^{-1} = \cos\th_{k+1}\,\G_{j,k+1} +
\sin\th_{k+1}\, \G_{j,k+2}\;,
\ee
one finds that \(U_j^{\;\;k+1}\) satisfies 
(\ref{lemma}) too. This finishes the proof of (\ref{lemma}) and thus 
the proof that (\ref{killing}) is the Killing spinor of AdS\(_{d+1}\).

\section{Asymptotics of AAdS spacetimes \label{proof}}

We obtain in this appendix the coefficients
$K_{(2n)ij}[g_{(0)}], n<d$ for AAdS. As mentioned in the main text,
it is sufficient to compute them for the exact $AdS$ solution.

Consider $AdS_{d+1}$ with boundary metric $g_{(0)}$ the standard
metric on $R \times S^{d-1}$. Then from (\ref{AdScoef}) we obtain,
\be
g_{(2)ab}=-\half g_{(0)ab}, \qquad g_{(2)tt}=\half g_{(0)tt},
\qquad g_{(4)ij} = \frac{1}{16} g_{(0)ij}.
\ee
The induced metric
and second fundamental form are given by
\bea
\g_{ij} &=& e^{2r} (g_{(0)ij} + e^{-2 r} g_{(2)ij} + e^{-4r} g_{(4)ij})
\nonumber \\
K_{ij}[\g] &=& \half \dot{\g}_{ij} = e^{2r} g_{(0)ij} - e^{-2r}
g_{(4)ij} \equiv K_{(0)ij}[\g] + \cdots + K_{(2n)ij}[\g] + \cdots \label{KAdS}
\eea
Recall that the coefficient $K_{(2n)ij}$ are local polynomials
of dimension $n$ of (covariant derivatives of the) curvature tensor of the induced metric.
For the case at hand, this implies that $K_{(2n)ij}$ is proportional to $R^n$, where
\be \label{g}
R = (d-1) (d-2) \g^{-1}, \qquad \g = e^{2r} \left(1-\frac{e^{-2 r}}{4}\right)^2
\ee
is the curvature scalar of $\g_{ij}$. Thus, the expansion in eigenfunctions of the dilation
operator is equivalent to an expansion in $\g^{-1}$,
\be
K_{ab}[\g]=\g_{ab} \sum_{n=0} \hat{k}_{(2n)} \g^{-2n}
\ee
Inserting this expression in (\ref{KAdS}) yields,
\be
\sum_{n=0} \hat{k}_{(2n)} e^{-2 n r} \left(1-\frac{e^{-2 r}}{4}\right)^{-2 n} =
(1 + \frac{e^{-2r}}{4})(1 - \frac{e^{-2r}}{4})^{-1}.
\ee
Expanding both sides around $r \to \infty$ and matching powers of $e^{-2 r}$ determines the
coefficients $\hat{k}_{(2n)}$. The first few are given in (\ref{k_coeff}).

We next turn to the expansion of the Witten spinor. As mentioned in the main 
text, we
take the Witten spinor to be a Killing spinor up to sufficiently high order, 
and the
Killing spinor is given by
\be \label{ksp_r}
\e_K(x,r) = e^{\frac{r}{2}} \hat{\e}_{(-\half)}(x) + e^{-\frac{r}{2}} 
\hat{\e}_{(\half)}(x).
\ee
To obtain the eigenfunctions of the dilatation operator we should express 
$\e_K(x,r)$
as a series in $\g^{-1}$,
\be \label{ksp_g}
e_K(x,r)=\sum_{m=0} \hat{\e}_{(-\half+m)}(x) \g^{-\frac{1}{2}(-\half+m)}.
\ee
Comparing (\ref{ksp_r}) and (\ref{ksp_g}) determines 
$\hat{\e}_{(-\half+m)}(x)$.
The first few are given in (\ref{e_coeff}).


\begin{thebibliography}{99}

\bibitem{AM} %2
A. Ashtekar and A. Magnon, ``Asymptotically anti-de Sitter space-times'',
Class. Quant. Grav. {\bf 1} (1984) L39-L44;
%\bibitem{AD} 
A.~Ashtekar and S.~Das,
  ``Asymptotically anti-de Sitter space-times: Conserved quantities,''
  Class.\ Quant.\ Grav.\  {\bf 17} (2000) L17
  [arXiv:hep-th/9911230].
  %%CITATION = HEP-TH 9911230;%%

%\cite{Henneaux:1984xu} %3
\bibitem{Henneaux:1984xu}
M.~Henneaux and C.~Teitelboim,
``Hamiltonian Treatment Of Asymptotically Anti-De Sitter Spaces,''
Phys.\ Lett.\ B {\bf 142} (1984) 355;
%%CITATION = PHLTA,B142,355;%%
``Asymptotically Anti-De Sitter Spaces,''
Commun.\ Math.\ Phys.\  {\bf 98}, 391 (1985).
%%CITATION = CMPHA,98,391;%%

\bibitem{FG} %1
C. Fefferman and C. Robin Graham, ``Conformal Invariants'', in
{\it Elie Cartan et les Math\'ematiques d'aujourd'hui} (Ast\'erisque, 1985)
95.


%\cite{Brown:1992br} %4
\bibitem{Brown:1992br}
  J.~D.~Brown and J.~W.~.~York,
  ``Quasilocal energy and conserved charges derived from the gravitational
  action,''
  Phys.\ Rev.\ D {\bf 47}, 1407 (1993).
  %%CITATION = PHRVA,D47,1407;%%

\bibitem{HS} %5
M.~Henningson and K.~Skenderis,
``The holographic Weyl anomaly,''
JHEP {\bf 9807} (1998) 023
[hep-th/9806087];
%%CITATION = HEP-TH 9806087;%%
M.~Henningson and K.~Skenderis,
``Holography and the Weyl anomaly,''
Fortsch.\ Phys.\  {\bf 48} (2000) 125
[hep-th/9812032].
%%CITATION = HEP-TH 9812032;%%

\bibitem{BK} %6
  V.~Balasubramanian and P.~Kraus,
  ``A stress tensor for anti-de Sitter gravity,''
  Commun.\ Math.\ Phys.\  {\bf 208}, 413 (1999)
  [arXiv:hep-th/9902121].
  %%CITATION = HEP-TH 9902121;%%

%\cite{Kraus:1999di}
\bibitem{Kraus:1999di}
P.~Kraus, F.~Larsen and R.~Siebelink,
``The gravitational action in asymptotically AdS and flat spacetimes,''
Nucl.\ Phys.\ B {\bf 563} (1999) 259
[hep-th/9906127].

\bibitem{dHSS} %7
S.~de Haro, S.~N.~Solodukhin and K.~Skenderis,
``Holographic reconstruction of spacetime and renormalization in the
 AdS/CFT correspondence,''
Commun.\ Math.\ Phys.\  {\bf 217} (2001) 595
[hep-th/0002230].
%%CITATION = HEP-TH 0002230;%%

\bibitem{Skenderis:2000in} %8 
K.~Skenderis,
``Asymptotically anti-de Sitter spacetimes and their stress energy  tensor,''
Int.\ J.\ Mod.\ Phys.\ A {\bf 16}, 740 (2001)
[arXiv:hep-th/0010138].
%%CITATION = HEP-TH 0010138;%%

\bibitem{howtogo}
M.~Bianchi, D.~Z.~Freedman and K.~Skenderis,
``How to go with an RG flow,''
JHEP {\bf 0108} (2001) 041
[arXiv:hep-th/0105276].
%%CITATION = HEP-TH 0105276;%%

\bibitem{holren}
M.~Bianchi, D.~Z.~Freedman and K.~Skenderis,
``Holographic renormalization,''
Nucl.\ Phys.\ B {\bf 631} (2002) 159
[arXiv:hep-th/0112119].
%%CITATION = HEP-TH 0112119;%%

%\cite{Martelli:2002sp}
\bibitem{Martelli:2002sp}
D.~Martelli and W.~Muck,
``Holographic renormalization and Ward identities with the
Hamilton-Jacobi method,''
Nucl.\ Phys.\ B {\bf 654} (2003) 248
[arXiv:hep-th/0205061];
%%CITATION = HEP-TH 0205061;%%

%\cite{Skenderis:2002wp}
\bibitem{Skenderis:2002wp}
K.~Skenderis,
``Lecture notes on holographic renormalization,''
Class.\ Quant.\ Grav.\  {\bf 19} (2002) 5849
[hep-th/0209067].
%%CITATION = HEP-TH 0209067;%%


 %\cite{Papadimitriou:2004ap} %21
\bibitem{PS1}
  I.~Papadimitriou and K.~Skenderis,
  ``AdS/CFT correspondence and geometry,''
  arXiv:hep-th/0404176.
  %%CITATION = HEP-TH 0404176;%%

%\cite{Papadimitriou:2004rz} %20
\bibitem{PS2}
  I.~Papadimitriou and K.~Skenderis,
  ``Correlation functions in holographic RG flows,''
  JHEP {\bf 0410} (2004) 075
  [arXiv:hep-th/0407071].
  %%CITATION = HEP-TH 0407071;%%

\bibitem{PS3} %12
  I.~Papadimitriou and K.~Skenderis,
  ``Thermodynamics of Asymptotically locally AdS spacetimes,'' 
  JHEP {\bf 08} (2005) 004
hep-th/0505190.
%%CITATION = HEP-TH 0505190;%%


%\cite{Abbott:1981ff} %9
\bibitem{Abbott:1981ff}
  L.~F.~Abbott and S.~Deser,
  ``Stability Of Gravity With A Cosmological Constant,''
  Nucl.\ Phys.\ B {\bf 195} (1982) 76.
  %%CITATION = NUPHA,B195,76;%%

%\cite{Hawking:1995fd} %10 
\bibitem{Hawking:1995fd}
  S.~W.~Hawking and G.~T.~Horowitz,
  ``The Gravitational Hamiltonian, action, entropy and surface terms,''
  Class.\ Quant.\ Grav.\  {\bf 13} (1996) 1487
  [arXiv:gr-qc/9501014].
  %%CITATION = GR-QC 9501014;%%

%\cite{Hollands:2005wt} %11 
\bibitem{Hollands:2005wt}
  S.~Hollands, A.~Ishibashi and D.~Marolf,
  ``Comparison between various notions of conserved charges in asymptotically
  AdS-spacetimes,''
  arXiv:hep-th/0503045.
  %%CITATION = HEP-TH 0503045;%%


\bibitem{Wald&Zoupas} %13
  R.~M.~Wald and A.~Zoupas,
  ``A General Definition of "Conserved Quantities" in General Relativity and
  Other Theories of Gravity,''
  Phys.\ Rev.\ D {\bf 61}, 084027 (2000)
  [arXiv:gr-qc/9911095] and references therein.
  %%CITATION = GR-QC 9911095;%%

\bibitem{GHW} %14 
  G.~W.~Gibbons, C.~M.~Hull and N.~P.~Warner,
  ``The Stability Of Gauged Supergravity,''
  Nucl.\ Phys.\ B {\bf 218} (1983) 173.
  %%CITATION = NUPHA,B218,173;%%

%\cite{Witten:1981mf} %15
\bibitem{Witten:1981mf}
  E.~Witten,
  ``A Simple Proof Of The Positive Energy Theorem,''
  Commun.\ Math.\ Phys.\  {\bf 80}, 381 (1981).
  %%CITATION = CMPHA,80,381;%%

%\cite{Witten:1981gj} %16
\bibitem{Witten:1981gj}
  E.~Witten,
  ``Instability Of The Kaluza-Klein Vacuum,''
  Nucl.\ Phys.\ B {\bf 195}, 481 (1982).
  %%CITATION = NUPHA,B195,481;%%

\bibitem{lebrun} %17
C. LeBrun, ``Counter-Examples to the Generaized Positive Action Conjecture'',
Commun. Math. Phys. {\bf 118} (1988) 591-596.

%\cite{Horowitz:1998ha} %30
\bibitem{Horowitz:1998ha}
  G.~T.~Horowitz and R.~C.~Myers,
  ``The AdS/CFT correspondence and a new positive energy conjecture for
  general relativity,''
  Phys.\ Rev.\ D {\bf 59}, 026005 (1999)
  [arXiv:hep-th/9808079].
  %%CITATION = HEP-TH 9808079;%%

\bibitem{Graham} C.R. Graham, ``Volume and Area Renormalizations for
Conformally Compact Einstein Metrics'', math.DG/9909042.

\bibitem{anderson} %19
  M.~T.~Anderson,
  ``Geometric aspects of the AdS/CFT correspondence,''
  arXiv:hep-th/0403087.
  %%CITATION = HEP-TH 0403087;%%


%\cite{Skenderis:1999nb} %18
\bibitem{Skenderis:1999nb}
K.~Skenderis and S.~N.~Solodukhin,
``Quantum effective action from the AdS/CFT correspondence,''
Phys.\ Lett.\ B {\bf 472} (2000) 316
[arXiv:hep-th/9910023].
%%CITATION = HEP-TH 9910023;%%


\bibitem{PS} %22
A.~Petkou and K.~Skenderis,
``A non-renormalization theorem for conformal anomalies,''
Nucl.\ Phys.\ B {\bf 561} (1999) 100
[arXiv:hep-th/9906030].
%%CITATION = HEP-TH 9906030;%%
  
\bibitem{Nester} %23
 J. Nester, ``A new gravitational expression with
a simple positivity proof'', Phys. Lett. {\bf 83A}, 241 (1981).

%\cite{Freedman:2003ax} %25
\bibitem{Freedman:2003ax}
  D.~Z.~Freedman, C.~Nunez, M.~Schnabl and K.~Skenderis,
  ``Fake supergravity and domain wall stability,''
  Phys.\ Rev.\ D {\bf 69}, 104027 (2004)
  [arXiv:hep-th/0312055].
  %%CITATION = HEP-TH 0312055;%%

\bibitem{GHHP} %24 
  G.~W.~Gibbons, S.~W.~Hawking, G.~T.~Horowitz and M.~J.~Perry,
  ``Positive Mass Theorems For Black Holes,''
  Commun.\ Math.\ Phys.\  {\bf 88}, 295 (1983).
  %%CITATION = CMPHA,88,295;%%


%\cite{Davis:1986da} %27
\bibitem{Davis:1986da}
  S.~Davis,
  ``Definition Of Conserved Quantities In Asymptotically Anti-De Sitter
  Space-Times,''
  Phys.\ Lett.\ B {\bf 166} (1986) 127.
  %%CITATION = PHLTA,B166,127;%%

%\cite{Cappelli:1988vw} %26
\bibitem{Cappelli:1988vw}
  A.~Cappelli and A.~Coste,
  ``On The Stress Tensor Of Conformal Field Theories In Higher Dimensions,''
  Nucl.\ Phys.\ B {\bf 314}, 707 (1989).
  %%CITATION = NUPHA,B314,707;%%
  

%\cite{Izquierdo:1994jz}
\bibitem{Izquierdo:1994jz}
  J.~M.~Izquierdo and P.~K.~Townsend,
  ``Supersymmetric space-times in (2+1) adS supergravity models,''
  Class.\ Quant.\ Grav.\  {\bf 12} (1995) 895
  [arXiv:gr-qc/9501018].
  %%CITATION = GR-QC 9501018;%%


\bibitem{BTZ} %28
  M.~Banados, C.~Teitelboim and J.~Zanelli,
  ``The Black hole in three-dimensional space-time,''
  Phys.\ Rev.\ Lett.\  {\bf 69}, 1849 (1992)
  [arXiv:hep-th/9204099];
  %%CITATION = HEP-TH 9204099;%%
M.~Banados, M.~Henneaux, C.~Teitelboim and J.~Zanelli,
  ``Geometry of the (2+1) black hole,''
  Phys.\ Rev.\ D {\bf 48}, 1506 (1993)
  [arXiv:gr-qc/9302012].
  %%CITATION = GR-QC 9302012;%%

%\cite{Coussaert:1993jp} %29
\bibitem{Coussaert:1993jp}
  O.~Coussaert and M.~Henneaux,
  ``Supersymmetry of the (2+1) black holes,''
  Phys.\ Rev.\ Lett.\  {\bf 72}, 183 (1994)
  [arXiv:hep-th/9310194].
  %%CITATION = HEP-TH 9310194;%%



%\cite{Constable:1999gb} %31
\bibitem{Constable:1999gb}
  N.~R.~Constable and R.~C.~Myers,
  ``Spin-two glueballs, positive energy theorems and the AdS/CFT
  correspondence,''
  JHEP {\bf 9910}, 037 (1999)
  [arXiv:hep-th/9908175].
  %%CITATION = HEP-TH 9908175;%%


%\cite{Galloway:2001uv} %32
\bibitem{Galloway:2001uv}
  G.~J.~Galloway, S.~Surya and E.~Woolgar,
  ``A uniqueness theorem for the adS soliton,''
  Phys.\ Rev.\ Lett.\  {\bf 88}, 101102 (2002)
  [arXiv:hep-th/0108170];
  %%CITATION = HEP-TH 0108170;%%
  ``On the geometry and mass of static, asymptotically AdS spacetimes, and  the
  uniqueness of the AdS soliton,''
  Commun.\ Math.\ Phys.\  {\bf 241}, 1 (2003)
  [arXiv:hep-th/0204081].
  %%CITATION = HEP-TH 0204081;%%

  \bibitem{Townsend:1984iu} %33
  P.~K.~Townsend,
  ``Positive Energy And The Scalar Potential In Higher Dimensional
  (Super)Gravity Theories,''
  Phys.\ Lett.\ B {\bf 148}, 55 (1984).
  %%CITATION = PHLTA,B148,55;%%




\bibitem{Skenderis:1999mm} %34
  K.~Skenderis and P.~K.~Townsend,
  ``Gravitational stability and renormalization-group flow,''
  Phys.\ Lett.\ B {\bf 468}, 46 (1999)
  [arXiv:hep-th/9909070].
  %%CITATION = HEP-TH 9909070;%%
  
  
  
  
  %\cite{Lu:1998nu} %35
\bibitem{Lu:1998nu}
  H.~Lu, C.~N.~Pope and J.~Rahmfeld,
  ''A construction of Killing spinors on $S^n$,''
  J.\ Math.\ Phys.\  {\bf 40} (1999) 4518
  [arXiv:hep-th/9805151].
  %%CITATION = HEP-TH 9805151;%%

\end{thebibliography}
\end{document}